\def\umass{1}
\def\upc{2}
\def\ieec{3}
\def\flash{4}
\def\aa{5}
\def\exeter{6}
\def\bethe{7}
\def\fub{8}

\newcommand{\sneia}{SNe~Ia}

\documentclass[apjl]{emulateapj}

\usepackage{amsmath}
\usepackage{ulem} 
\usepackage{subfigure} 
\usepackage{bm} 

\begin{document}

\submitted{Accepted in ApJ}

\title{The Post-Merger Magnetized Evolution of  White Dwarf Binaries: The Double-Degenerate Channel of Sub-Chandrasekhar  Type Ia Supernovae and the Formation of Magnetized White Dwarfs}

\author{
Suoqing~Ji,\altaffilmark{\umass}
Robert~T.~Fisher,\altaffilmark{\umass}
Enrique~Garc\'ia-Berro,\altaffilmark{\upc,\ieec}
Petros~Tzeferacos,\altaffilmark{\flash,\aa}
George~Jordan,\altaffilmark{\flash,\aa}
Dongwook~Lee,\altaffilmark{\flash,\aa}
Pablo~Lor\'en-Aguilar,\altaffilmark{\exeter}
Pascal~Cremer,\altaffilmark{\bethe}
Jan~Behrends \altaffilmark{\fub}
}

\altaffiltext{\umass}{University of Massachusetts Dartmouth, Department of Physics, 285 Old Westport Road, North Dartmouth, 02740.}

\altaffiltext{\upc}{Departament de F\'{i}sica Aplicada, Universitat Polit\`{e}cnica de Catalunya, c/Esteve Terrades, 5, 08860 Castelldefels, Spain.}

\altaffiltext{\ieec}{Institut d'Estudis Espacials de Catalunya, Ed. Nexus-201, c/Gran Capit\`a 2-4, 08034 Barcelona, Spain.}

\altaffiltext{\flash}{Center for Astrophysical Thermonuclear Flashes, The University of Chicago, Chicago, IL 60637.}

\altaffiltext{\aa}{Department of Astronomy and Astrophysics, The University of Chicago, Chicago, IL 60637.}

\altaffiltext{\exeter}{School of Physics, University of Exeter, Stocker Road, Exeter EX4 4QL.}

\altaffiltext{\bethe}{Bethe Center for Theoretical Physics, Universit\"{a}t  Bonn, Nussallee 12, 53115 Bonn, Germany.}

\altaffiltext{\fub}{Fachbereich Physik, Freie Universit\"{a}t Berlin,  Arnimallee 14, 14195 Berlin, Germany.}

\begin{abstract}
Type Ia supernovae (SNe Ia) play a crucial role as standardizable cosmological candles, though the nature of their progenitors is a subject of active investigation. Recent observational and theoretical work has pointed to  merging white dwarf binaries, referred to as the double-degenerate channel, as the possible progenitor systems for some SNe~Ia. Additionally, recent theoretical work suggests that mergers which fail to detonate may produce magnetized, rapidly-rotating white dwarfs. In this paper, we present the first multidimensional simulations of the post-merger evolution of white dwarf binaries to include the effect of the magnetic field.  In these systems, the two white dwarfs complete a final merger on a dynamical timescale, and are tidally disrupted, producing a rapidly-rotating white dwarf merger surrounded by a hot corona and a thick, differentially-rotating disk. The disk  is strongly susceptible to the magnetorotational instability (MRI), and we demonstrate that this leads to the rapid growth of an initially dynamically weak magnetic field in the disk,  the spin-down of the white dwarf merger, and to the subsequent central ignition of the white dwarf merger. Additionally,  these magnetized models exhibit new features not present in prior hydrodynamic studies of white dwarf mergers, including the development of MRI turbulence in the hot disk, magnetized outflows carrying a significant fraction of the disk mass, and the magnetization of the white dwarf merger to field strengths $\sim 2 \times 10^8$~G. We discuss the impact of our findings on the origins, circumstellar media, and observed properties of SNe~Ia and magnetized white dwarfs.


\end{abstract}


\section{Introduction}

Type Ia supernovae (\sneia) are among the most energetic explosions in the known universe, releasing 10$^{51}$~erg of kinetic energy  and synthesizing $0.7\, M_{\sun}$ of radioactive $^{56}$Ni in the ejecta of a typical brightness explosion.  The discovery of the Phillips relation \citep {phillips93} enabled the use of Type Ia supernova (SN Ia) as standardizable cosmological candles, and has ushered in a new era of astronomy leading to the discovery of the acceleration of the universe \citep {riessetal98, perlmutteretal99}.

The nature of the Type Ia progenitors, as well as their precise explosion mechanism, remains a subject of active investigation, both observationally as well as theoretically.  Observational progress to determine the nature of the SNe~Ia progenitors, as well as their underlying explosion mechanism, has accelerated in recent years, with a series of projects, including the Palomar Transient Factory (PTF), Large Synoptic Survey Telescope, Pan-STARRS, and the Dark Energy Survey, all coming online. This progress culminated in the discovery of 2011fe in M101 by PTF on August 24, 2011 \citep {nugentetal11}.  At a distance of 6.4 Mpc, 2011fe is the nearest SNe Ia detected in the last 25 years, and has proven to be the kind of SNe~Ia exemplar system that SN~1987A has been for SNe~II.  PTF captured 2011fe within 11 hours of the explosion, making it the earliest SN~Ia ever detected, and opening the gates to prompt multi-waveband follow-up observations in the radio, optical, UV, and X-ray bands.   The first weeks of multi-wavelength follow-up observations have directly confirmed for the first time that the primary object is a carbon-oxygen white dwarf, and have placed tight constraints on the progenitor system to 2011fe, ruling out red giant as well as Roche-lobe overflowing main sequence companions \citep {nugentetal11, brownetal12, horeshetal12, lietal11, bloometal12}. 

Additionally, determinations of the SNe~Ia delay time distribution (DTD) generally follow  a $t^{-1}$ power-law, consistent with expectations for  double-degenerate (DD) systems \citep {galyammaoz04, totanietal08, maozbadenes10, maozetal10}. A related, but independent model of the supernovae rate based upon a two-component model accounting for both a prompt and delayed component also supports the existence of a delayed, DD channel \citep {scannapiecobildsten05, raskinetal09b}. Moreover, the search for both point sources  in pre-explosion archival data  and ``ex-companions'' in SNe~Ia remnants have so far yielded no definitive candidates, with the possible contested example of  Tycho's remnant \citep {maozmannucci08, gonzalezhernandezetal12, schaeferpagnotta12, edwardsetal12, kerzendorfetal12}.  Lastly, the search for hydrogen in the nebular spectra of remnants places fairly tight constraints on the amount of hydrogen ($\la 10^{-2}\, M_{\sun}$) than can be stripped from a companion \citep {leonard07}. It may be possible to understand both the absence of ex-companions and nebular hydrogen in the context of the single-degenerate channel as due to the time delay of the spin-down of the white dwarf after accretion from its companion ceases \citep {distefanokilic12}.   However, the weight of the observational evidence strongly suggests the viability of the DD channel model as the progenitors for some, if not the majority, of  SNe~Ia events.  

The key conceptual challenge faced by the DD  channel for SNe~Ia is to explain how these models yield a thermonuclear runaway, as opposed to an accretion-induced collapse (AIC) to a neutron star. Early spherically-symmetric models, based on Eddington accretion rates onto the white dwarf merger, suggested that double degenerate mergers will ignite a carbon-burning deflagration wave which propagates inward to the core of a $1.0\, M_{\sun}$ 50/50 carbon-oxygen white dwarf in $\sim 2 \times 10^4$~yr, comparable to the thermal timescale of the white dwarf merger, resulting in a AIC \citep {saionomoto85, saionomoto98, saionomoto04}. An AIC may be avoided, however, if the actual evolution within the rotating merger and fully multi-dimensional, magnetized accretion disk differs significantly from the one-dimensional models, and can ignite a detonation on a timescale much shorter than the inward carbon deflagration timescale \citep {nomotoiben85}. An additional crucial difference highlighted by more recent numerical simulations deals with the thermal structure of the white dwarf merger itself  \citep {mochkovitchlivio89, mochkovitchlivio90, rasioshapiro95, segretainetal97, guerreroetal04, dsouzaetal06, yoonetal07, motletal07, pakmoretal10, danetal11, zhuetal11}.  These simulations demonstrate that the compressional work produced during the final tidal disruption of the white dwarf binary results in a hot ($\sim 10^8$~K) rotating white dwarf merger, quite unlike the cold ($10^7$~K) isothermal white dwarfs typically taken as a starting point in earlier one-dimensional studies. 

Furthermore, recent work suggests that the outcome of white dwarf mergers may not always be either a SNe~Ia or an AIC, but could also result in a high-field magnetic white dwarf (HFMWDs). HFMWDs have magnetic fields in excess of $10^6$~G and up to $10^9$~G. Very few of these white dwarfs belong to a non-interacting binary system, and moreover they are more massive than average --- see, for instance, \cite{Kawka}. All these characteristics point towards a binary origin of these white dwarfs. Although long-suspected \citep {wickramasingheferrario00}, it has only recently been shown that if the white dwarf merger fails either to detonate into an SNe~Ia or collapse down into an AIC, it will result in a magnetized, rapidly-rotating white dwarf \citep {garciaberroetal12}. Whether the magnetic white dwarf rotates rapidly or not depends on the relative orientation of the magnetic and rotational axes, as well as on the efficiency of the various braking mechanisms. 

The evolution of the white dwarf merger subsequent to the coalescence of the initial binary system remains a subject of active investigation. Numerical models have begun to relax assumptions of earlier work by modeling the accretion of the hot thick accretion disk onto the white dwarf, either by including a prescription for the accretion process and the spin-down of the merger  \citep {yoonetal07}, or by employing a Shakura-Sunyaev turbulent viscosity \citep {shenetal12, schwabetal12}.   Other researchers have investigated the violent mergers of super-Chandrasekhar mass white dwarf systems \citep {pakmoretal10, pakmoretal11} and collisions \citep {raskinetal09a}. While violent mergers and collisions of white dwarfs are found by some groups to lead to detonations, these detonations may be sensitive to the initial conditions of the white dwarf merger \citep {motletal07, danetal11}. Because the detonations are not fully-resolved in large-scale multidimensional simulations, detonations must be initiated by the choice of a suitable criterion, which is an additional issue in determining whether these systems robustly detonate -- e.g.	 \citet {seitenzahletal09a}.

In contrast, sub-Chandrasekhar mergers are more prevalent than super-Chandrasekhar mergers in nature.  Recently, van Kerjwijk and colleagues have re-invigorated the examination of  whether the accretion of the disk  may give rise to a detonation, {\it even for sub-Chandrasekhar mergers} \citep {vankerkwijketal10, zhuetal11, vankerkwijk12}. Specifically, beginning with a near-equal mass binary with two  $0.6\, M_{\sun}$ carbon-oxygen white dwarfs, which both subsequently tidally disrupt and merge,  van Kerkwijk et al suggest that the accretion of the thick, turbulent disk surrounding the white dwarf merger results in the compressional heating of the degenerate material in the white dwarf over a viscous timescale, which in turn leads to a detonation.  The model explains how a realistic multidimensional DD merger might produce a SNe~Ia instead of an AIC. Equally interesting is the possibility that sub-Chandrasekhar DD mergers may help to bring observed and predicted SNe rates in closer agreement \citep {ruiteretal09, badenesmaoz12}. Furthermore, detonations of sub-Chandrasekhar progenitors naturally produce nucleosynthetic yields and luminosities closely in line with observation,  and thereby sidestep the long-standing problem of pre-expansion required during the deflagration phase of near-Chandrasekhar mass white dwarf progenitors encountered in the single-degenerate channel \citep {simetal10}.

Despite these advances,  all previous numerical simulations to-date on double-degenerates have treated the influence of  magnetic field only approximately through the use of a Shukura-Sunyaev $\alpha$ prescription, or neglected its contribution entirely. Furthermore, apart from recent work by \citet {romanovaetal11, romanovaetal12} and \citet {siegaletal13}, which focused upon the inner accretion disk surrounding a non-burning star, and a hypermassive neutron star, respectively, all work done on the magnetorotational instability (MRI) to date have generally lacked a central stellar object.  Consequently, fundamental questions involving the influence of the magnetic field upon the outcome of the WD merger and its connection to SNe~Ia remain unresolved.  These questions include:  What is the structure of the magnetic field in the white dwarf merger, disk, and corona --- ordered or disordered? If ordered magnetic fields are present, are they capable of collimating outflows? Are significant amounts of mass outfluxed from the binary system? How does the magnetic field influence the spin of the merger? What is the influence of the field upon nuclear burning? Under what conditions may we generally expect mergers to produce stable HFMWDs, SNe~Ia, or accretion-induced collapses? In this paper, we present the first set of numerical simulations of the post-merger evolution of double-degenerates to include the effect of the magnetic field, and in so doing, to begin to address the role of the magnetic field upon each of these fundamental questions. 


\section{Simulations of the Double-Degenerate Model\label{sec:DD-models}}
\subsection{Simulation Setup\label{subsec:simsetup}}
 The astrophysical fluid framework code FLASH has previously been used to simulate single-degenerate models of \sneia\ in both 2D cylindrical and 3D Cartesian geometry in a wide range of studies \citep{ townsleyetal07, jordanetal08, meakinetal09, faltaetal11, jordanetal12b, jordanetal12a}.  Here we build upon and extend this body of work to simulate the double-degenerate channel. Our double-degenerate models take as an initial condition the endpoint of a $0.6\, M_{\sun} + 0.6\, M_{\sun}$ 40/60 carbon-oxygen white dwarf merger from the 3D SPH simulations of \citet {lorenaguilaretal09}. The stars  are modeled using $4 \times 10^5$ particles. The initial condition corresponds to non-sychronously rotating white dwarfs on a circular orbit with an orbital separation  of  $\simeq   0.03\,  R_{\sun}$, which immediately leads to an unstable  mass  transfer  episode lasting for $\sim 500$~s. The  final  configuration  of  the  merger  corresponds  to  a  solidly-rotating central compact object ($\Omega \simeq 0.3$~s$^{-1}$) surrounded by a thick  accretion  disk extending  up  to  a  distance of $\simeq 0.07\, R_{\sun}$. No noticeable nuclear processing was found during the merger process.


 Unlike    unequal-mass    and synchronously-rotating merger   events,   the    remnant     of  the irrotational $0.6 + 0.6\, M_{\sun}$ case   presents   a   temperature  peak  ($\simeq 6 \times 10^8$~K) at  the center of the   central  compact object,   with   a   rapid drop    towards     the     outer    parts    of  the   disk.  Whether or not the initial white dwarf binary is synchronously-rotating depends upon the efficiency of tidal dissipation, which remains an open question. \citet {segretainetal97} argue that the timescale for gravitational wave emission is much longer than the orbital timescale for dynamically-unstable systems, leading to non-synchronous rotation of the white dwarfs. Later work by some groups find resonant tidal dissipation in the final merger process to be highly-efficient, thereby producing synchronous binaries \citep {burkartetal12}, while other groups find that a degree of asynchronicity persists even to short periods \citep {fullerlai13}.  Asynchronously-rotating systems generally lead to more violent mergers and hotter initial conditions  -- e.g., \citet {pakmoretal10} and \citet {zhuetal11}, while synchronously-rotating systems result in less violent mergers with lower temperatures -- e.g. \citet {danetal11}. Because the initial temperature profile plays a crucial role in determining the nuclear burning within the white dwarf merger, the possible influence of initial synchronous rotation and tidal heating of the white dwarf binary are important assumptions of our models which must be borne in mind. In particular, we expect the majority of our conclusions regarding the development of the magnetic field to be robust, though  these tidal effects will influence the temperature profile of the white dwarf merger, and  may impact our conclusions with regard to nuclear ignition.
 
 We utilize the SPH smoothing kernel to interpolate the Lagrangian SPH data onto a 2D axisymmetric cylindrical coordinate ($r$, $z$) Eulerian mesh, averaging over azimuthal angle $\phi$, while retaining all three velocity components ($v_r$, $v_{\phi}$, $v_z$) --- see appendix for details.  Both the interpolation and the azimuthal angle-averaging are not guaranteed to conserve energy; however, we have confirmed that the initial total energy on the mesh is preserved  to within 1\% or better of the SPH value for all models presented here. We then advance this initial Eulerian initial condition in time using the adaptive mesh refinement (AMR) FLASH application framework \citep{dubey2009, fryxelletal00} using an initially-weak poloidal magnetic field to seed the growth of the MRI. 
 
We solve the fundamental governing equations of self-gravitating, inviscid ideal magnetohydrodynamics, which can be expressed as:
 
 \begin {equation}
{\partial \rho \over \partial t} + {\bm \nabla} \cdot (\rho {\bm v}  ) = 0 
 \end {equation}
 
 \begin {equation}
 {\partial \rho {\bm v} \over \partial t} + {\bm \nabla} \cdot (\rho {\bm v} {\bm v} - {\bm B} {\bm B}  ) + {\bm \nabla p_*} = \rho {\bm g}  
 \end {equation}
 
 \begin {equation}
  {\partial \rho E \over \partial t} + {\bm \nabla} \cdot \left [ \bm v (\rho E + p_*) - {\bm B} ({\bm v} \cdot {\bm B}) \right] = \rho {\bm g} \cdot {\bm v}
 \end {equation}
 
  \begin {equation}
\label {eqn:induction}
 {\partial {\bm B} \over \partial t} + {\bm \nabla} \cdot ({\bm v} {\bm B} - {\bm B} {\bm v}) = 0 
 \end {equation}
  Here, $\rho$ is mass density, ${\bm v}$ is the fluid velocity, ${\bm B}$ is the magnetic field, and ${\bm g}$ is the gravitational acceleration. $p_* = p + B^2 / (8 \pi)$ is the total pressure, including both gas pressure $p$ and magnetic pressure $B^2 / (8 \pi)$. $\rho E$ is the total energy density, $E = 1/2 \rho v^2 + \rho \epsilon +  B^2 / (8 \pi)$.   The inclusion of Poisson's equation for self-gravity:
 \begin {equation}
 {\bm \nabla}^2 \phi = 4 \pi G \rho,
 \end {equation}
where ${\bm g} = - {\bm \nabla} \phi$, as well as an equation of state close the system.  Equation~(\ref {eqn:induction}), the induction equation, preserves ${\bm \nabla} \cdot {\bm B} = 0$ if this is imposed as an initial condition. 

 \begin{deluxetable*}{p{0.8cm}cccccccccc}[t!]
\tabletypesize{\scriptsize}
\tablecaption{Simulation Properties\label{tab:sims}}
\tablewidth{0pt}
\tablehead{
\colhead{Run} &
\colhead{$N^2$ \tablenotemark{a}} &
\colhead{$\Delta x\, (\rm {km})$ \tablenotemark{b}} &
\colhead{$\beta$ \tablenotemark{c}} &
\colhead{$E_{\rm mag}$ (erg) \tablenotemark{d}} &
\colhead{$\langle \langle \alpha_m \rangle \rangle$} &
\colhead{$H / \Delta x$ \tablenotemark{f}} &
\colhead{max ($\lambda_c / \Delta x $) \tablenotemark{g}}   &
\colhead{{$t_{\rm run}$ (s)} \tablenotemark{h} } &
\colhead{$M_{\rm out}$ ($M_{\odot}$) \tablenotemark{i} } &
\colhead{max ($T$) (K) \tablenotemark{j} } }

\startdata
S-bh       &       $512^2$ & 256       &      2000    &  $2.79 \times 10^{46}$ & 0.020 & 18.7  & 76.3  & $2 \times 10^4$ &  0.0312 & $7.29 \times 10^8$\\
H-bh       &       $1024^2$ & 128       &     2000   &  $2.79 \times 10^{46}$ &  0.023 & 37.4  & 158.2  & $5 \times 10^3$ & 0.0389 &  $7.79 \times 10^8$\\
L-bh        &       $256^2$ & 512       &     2000      &  $2.79 \times 10^{46}$ &  0.033 & 9.4 & 38.3 &  $5 \times 10^3$ & 0.0308 & $6.83 \times 10^8$\\
S-bm       &       $512^2$ & 256       &     1000      & $5.58 \times 10^{46}$ & 0.036 & 18.7 & 107.9  &$10^4$ & 0.058 & $8.00 \times 10^8$ \\
S-bl        &        $512^2$ & 256       &     500        & $1.12 \times 10^{47}$ & 0.033 & 18.7 & 152.6 & $1.5 \times 10^4$ & 0.027 & $8.60 \times 10^8$\\
\enddata
\tablenotetext{a}{Number of grid cells per domain.}
\tablenotetext{b}{Linear spatial resolution.}
\tablenotetext{c}{Initial global average of $\beta$, the dimensionless ratio of gas pressure to magnetic pressure, over the entire spatial domain.}
\tablenotetext{d}{Initial total magnetic energy on the problem domain.}
\tablenotetext{e}{Spatially and time-averaged magnetic stress parameter $\alpha_m$. }
\tablenotetext{f}{Ratio of isothermal disk scale height $H$ to $\Delta x$, evaluated near middle of disk.}
\tablenotetext{g}{Initial maximum ratio of the effective resolution $\lambda_c / \Delta x $.}
\tablenotetext{h}{Maximum duration of run.}
\tablenotetext{i}{Total mass outfluxed from the domain, including both bound and unbound mass.}
\tablenotetext{j}{Maximum temperature reached over the duration of the run.}
\end{deluxetable*}

 The poloidal magnetic field is efficiently initialized in a divergence-free form by defining the toroidal component of the vector potential $\bm {A}$:
%
%
\begin{equation}
 A_{\phi} (r) = \begin{cases} B_0 (\rho - \rho_0)  f(r)  &\mbox{if } \rho > \rho_0 \\
0 & \mbox{if } \rho \le \rho_0. \end{cases} 
\end{equation}
This form is motivated by previous MRI studies \citep {hawley00}, with the inclusion of a filter function $f (r) =  \Delta\ {\rm tanh} \left[ \left(  r - r_0 \right) / \Delta  \right]^\alpha$,  chosen to localize the initial poloidal field to the disk, and to avoid initially strongly magnetizing the merger, the axial region near the $z$-axis, as well as low-density regions outside the disk.    Here we have chosen $r_0 = 10$~km, $\Delta = 1.5 \times 10^4$~km, $\rho_0 = 2$~g/cm$^3$, $\alpha = 9$,  and $B_0$ is an overall field strength factor chosen to ensure the magnetic pressure is everywhere weak compared to the gas pressure initially. 
The magnetic field is then straightforwardly defined at cell edges by finite-differencing the vector potential using $\bm {B} =  {\bm \nabla} \times \bm {A}$, and advanced using the unsplit ideal MHD solver in angular-momentum conserving form \citep {leedeane09, tzeferacosetal12, lee13}.  The Roe Riemann solver is employed, with piecewise parabolic (PPM) spatial reconstruction and a minmod slope limiter. The divergence-free prolongation of the magnetic field is done using an adapted implementation of the method of
\citet{lili04}.

 
Our simulations employ an equation of state that includes contributions from blackbody radiation, ions, and electrons of an arbitrary degree of degeneracy \citep{2000ApJS..126..501T}, along with an axisymmetric multipole treatment of gravity, with the series truncated after ten moments ($\ell = 10$). Nuclear burning is incorporated through the use of a simplified 13-species alpha-chain network, which includes the effect of neutrino cooling \citep {timmes99}.

\begin{figure*}[th!]
\begin{center}
  \subfigure[$\log\rho$]{
   \label{fig:logdens}
    \includegraphics[width=0.49 \textwidth]{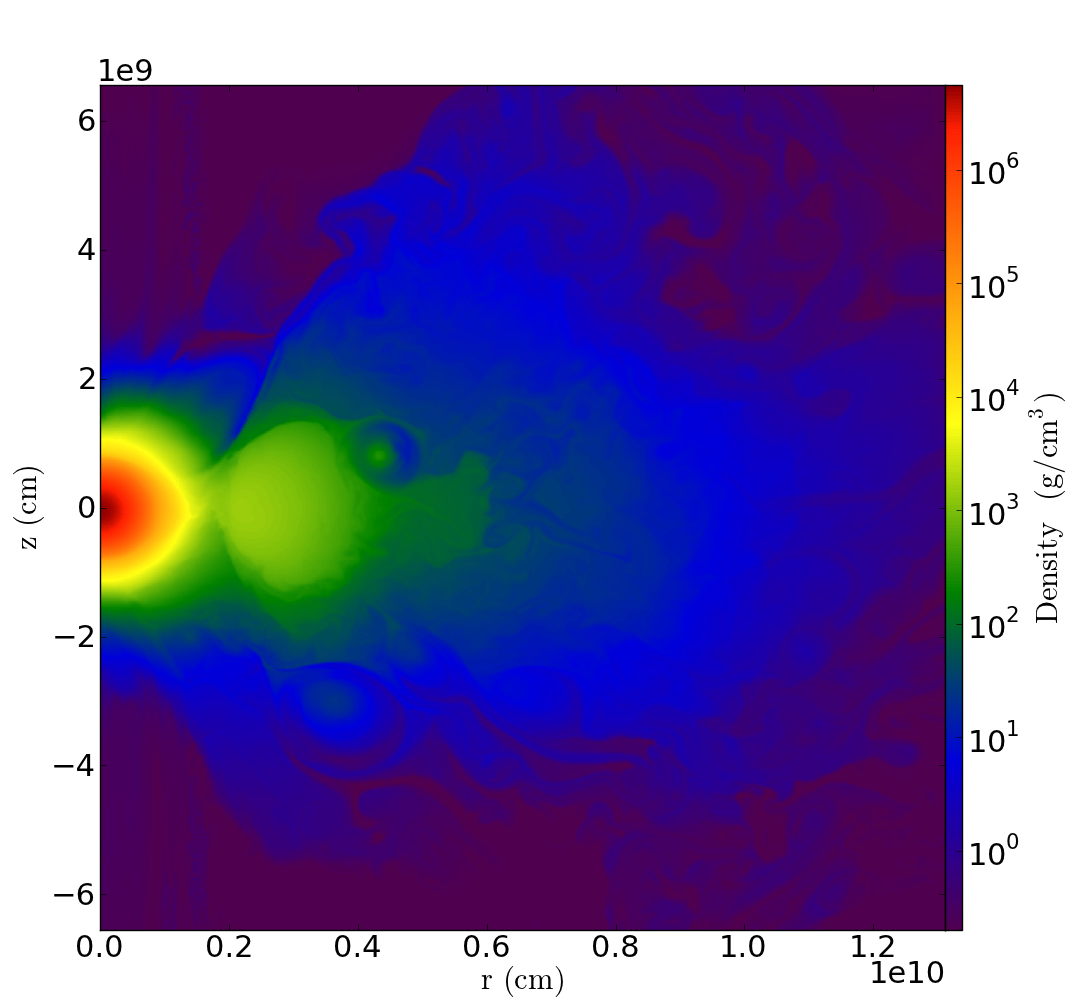}}
    \subfigure[$\log T$]{
    \label{fig:logtemp}
    \includegraphics[width=0.49 \textwidth]{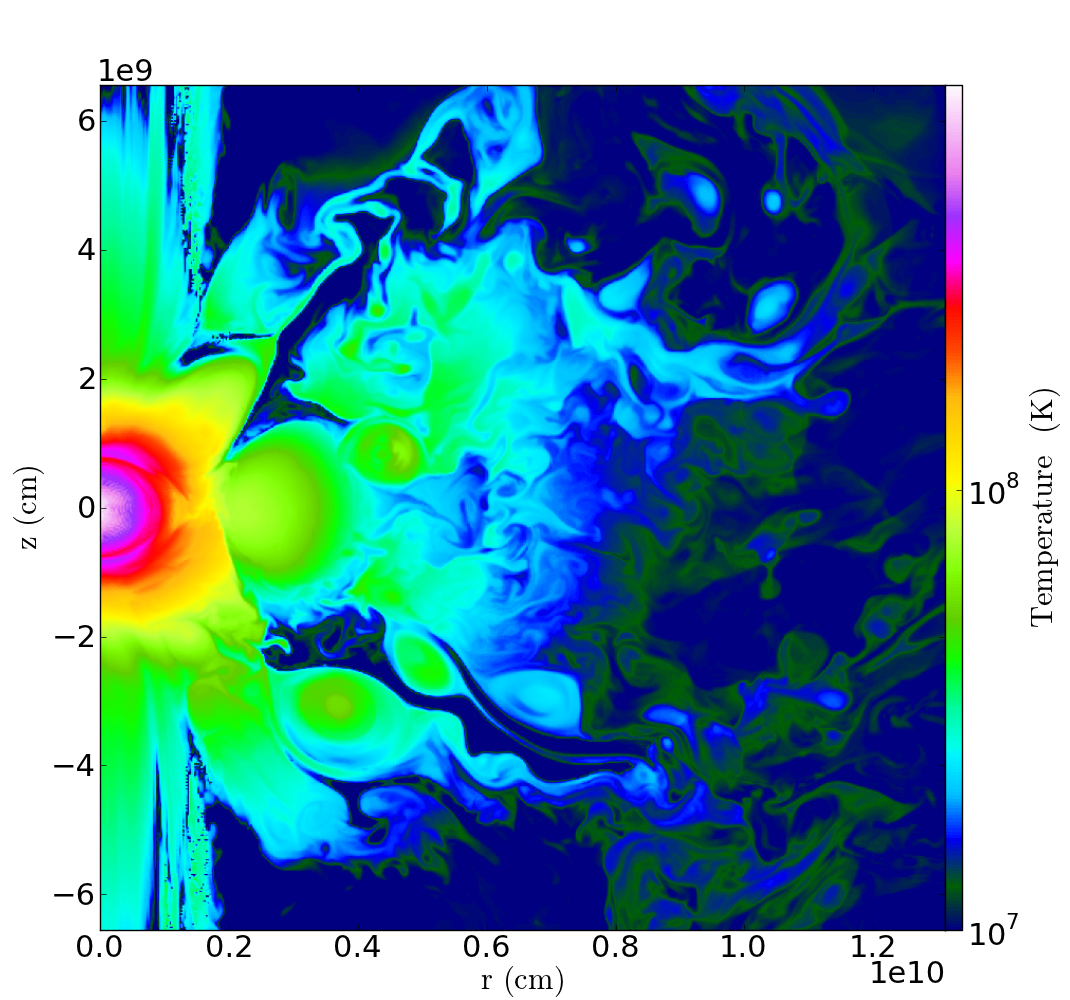}}
  \\
  \subfigure[Magnetic Field]{
    \label{fig:logmagz}
    \includegraphics[width=0.49 \textwidth]{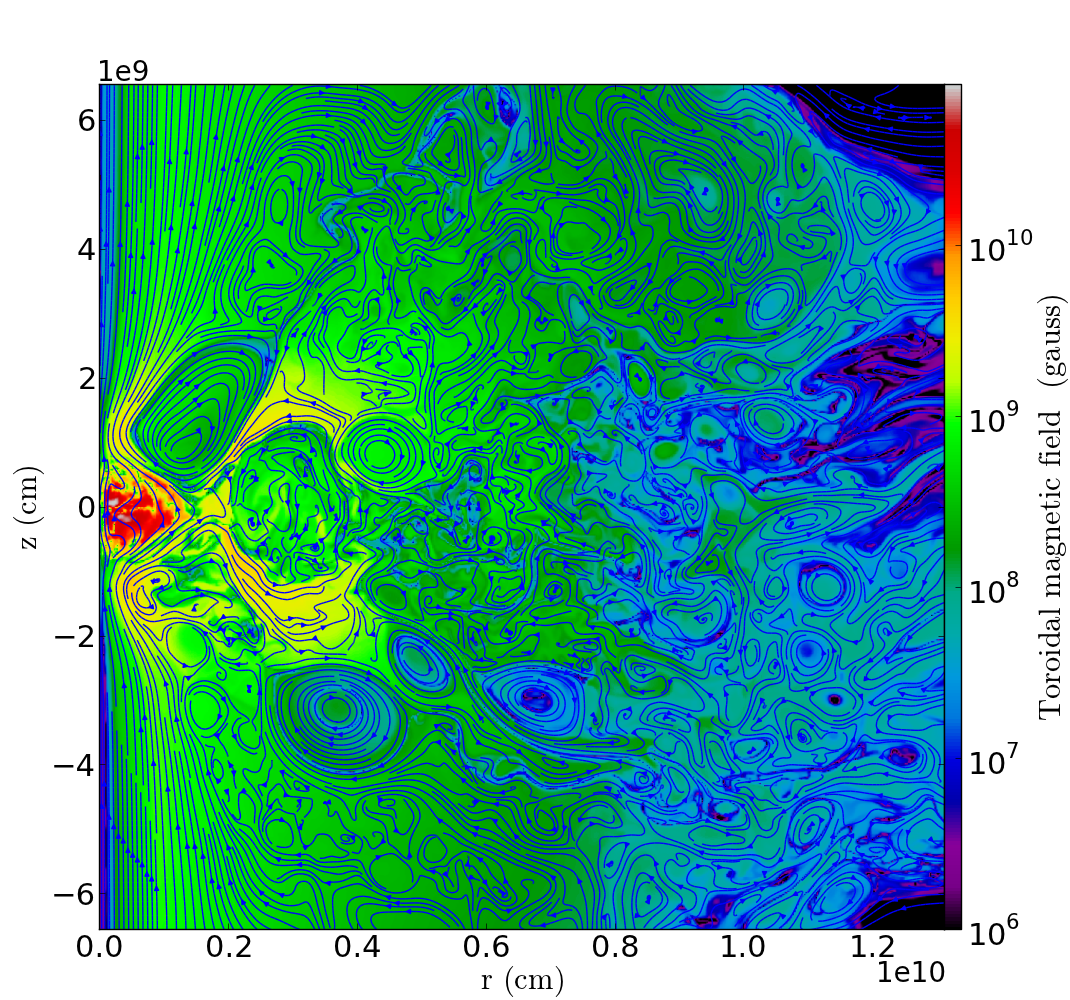}}
   \subfigure[$\log\beta$]{
    \label{fig:logbeta}
   \includegraphics[width=0.49 \textwidth]{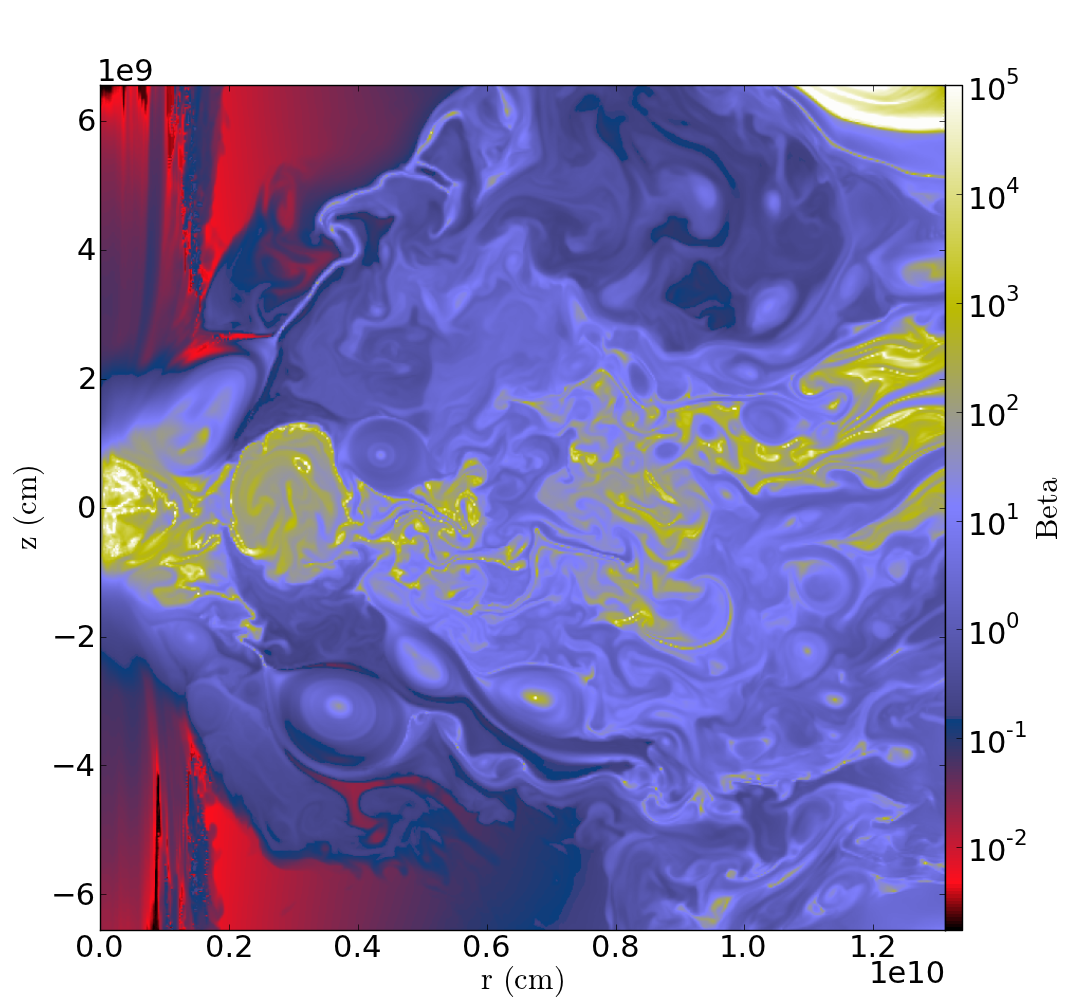}}
    
\caption{Four frames in the $r$-$z$ plane consisting of a) $\log\rho$, b) $\log T$, c) magnetic field, with lines of poloidal magnetic field in the $r$-$z$ plane superposed against a color raster plot of the toroidal field $B_\phi$, and  d) the ratio of gas pressure to magnetic pressure $\beta$ value. All four frames are taken at the midpoint of the model S-bh simulation at $t = 10^4$~s.}
\label{fig:panel}
\end{center}
\end{figure*}

We performed a series of simulations from the previously-described initial condition, varying both our choice of the initially-weak magnetic field as well as the spatial resolution. The full set of completed runs is shown in table~\ref {tab:sims}. All runs are performed on a domain   $r < 1.31 \times 10^{10}$~cm in the radial direction, and $-6.55 \times 10^9$~cm~$< z < 6.55 \times 10^9$~cm in the vertical, with diode boundary conditions, which strictly guarantee that no inflow occurs at the outer boundary. The white dwarf merger has a radius $\simeq 1.5 \times 10^4$~km, which is resolved with 60 cells in our standard model. The isothermal disk scale height $H = \sqrt {2} c_s  /\Omega \simeq 4.8 \times 10^8$~cm at a temperature of $5 \times 10^7$~K near the midpoint of the disk, where $\Omega \simeq 0.05$~s$^{-1}$.  At an inner disk radius of $2 \times 10^9$~cm, the Keplerian period around a $1.0\, M_{\sun}$ white dwarf merger is 51~s. The viscous accretion time, based upon a simple constant Shakura-Sunyaev $\alpha = 0.01$, is roughly $7 \times 10^3$~s.   Our standard model (designated ``S-bh'')  has a resolution of 256~km  and an initial value of the global ratio of gas pressure to magnetic pressure, defined to be the  dimensionless ratio $\beta = 8 \pi {\langle } P \rangle  / \langle B \rangle^2 = 2000$, where angle bracket quantities represent averages over space over the entire domain. While the local ratio of  gas to magnetic pressure varies throughout the spatial domain, the magnetic pressure is  initially significantly less than the initial total gas pressure in all models; its initial minimum value is 16.5 in model S-bh. We note that while our seed magnetic field is initially dynamically weak everywhere, and its initial magnitude in the white dwarf merger ($2.8 \times 10^5$~G)  is typical of field white dwarfs ($\leq 10^6$~G), its value in the disk is astrophysically large by this same comparison. Our choice is motivated by the requirement to accurately capture the dynamics of the MRI by well-resolving the fastest-growing MRI mode  $\lambda_c \simeq 6.49\ v_A / \Omega,$ where $v_A$ is the local Alfv\'en speed, and $\Omega$ the local rotational velocity  \citep {hawleygammiebalbus95}, on a computationally-tractable mesh size.  MRI simulations which do not initially resolve $\lambda_c$ also become unstable and reach magnetic field saturation, but take much longer to do so, since only a narrow band of all unstable modes, not including the fastest-growing mode, are captured on the mesh \citep {suzukiinutsuka09}. In reality, because the fastest-growing mode of the MRI always grows on a dynamical timescale $\sim \Omega^{-1}$, we expect that even an astrophysically-realistic magnetic field strength will reach saturation on a relatively short timescale in comparison to the viscous timescale.

Our additional models vary both the spatial resolution and the initial magnetic field strength, in order to test spatial convergence, as well as sensitivity to the resolution of $\lambda_c$. Our standard model S-bh maximally resolves $\lambda_c$ with approximately 76 cells per wavelength, and the disk scale height $H$ (evaluated near the midpoint of the disk) with roughly 19 cells. The vertical resolution of our standard model is therefore somewhat  less than 3D convergence studies of stratified shearing-box MRI simulations, which demonstrate that between 32 and 64 cells per scale height are required for convergence \citep {davisetal10}. Moreover, because both the disk scale height in a global disk geometry and the initial seed magnetic field vary, we do not necessarily resolve either the scale height or  the fastest-growing mode uniformly throughout the domain.  To address this issue, we examine our convergence of the peak magnetic field and magnetic stresses by varying our resolution explicitly, increasing and decreasing our standard model resolutions in runs H-bh and L-bh, respectively. Additionally, we also vary our initial magnetic field strength, increasing it in both models S-bm and S-bl, which have initial $\beta$ values of 1000 and 500, respectively.  Because $\lambda_c$ depends on the initial field strength, these model variations also vary the effective resolution (see table~\ref {tab:sims}, column f).  

 Our standard model S-bh has been advanced to $2 \times 10^4$~s, while other models run for varying durations. Our standard model duration is equivalent to  390 inner rotational periods, and several viscous accretion timescales. We output the state of the system in 10~s intervals. This extensive time series is sufficiently long to permit accurate time-averages over turbulent quantities. We experimented with both relaxed and ``cold-start'' initial conditions. In general, due to small differences in numerical schemes, the initial conditions mapped from SPH lead to a mapped initial condition on the Eulerian mesh with small spurious radial oscillations in the disk and merger \citep {zingaleetal02}. In the relaxed cases, a damping term in the momentum equation drove the system to a hydrodynamic equilibrium state, eliminating the radial oscillations over a few dynamical times, prior to the introduction of the seed magnetic field. In contrast, cold starts simply allowed the system to evolve from $t = 0$ with the seed magnetic field.   Due to the rapid growth of the MRI, there were relatively small changes in the outcome between these initialization procedures. The results presented here are all cold starts.

\subsection{Simulation Results\label{subsec:simresults}}

\subsubsection{Magnetic Structure of White Dwarf Merger and Disk}

A snapshot depicting the logarithm of the density taken from the midpoint of our standard model, at $t = 10^4$~s,  is shown in figure~\ref {fig:logdens}, whereas the distribution of temperatures is shown in figure~\ref {fig:logtemp}. The rotating, hot white dwarf merger is surrounded by a differentially-rotating, thick accretion disk. On the bottom panels, figures~\ref{fig:logmagz} and \ref {fig:logbeta}, we reveal the magnetic structure of the merger and the disk. This is done plotting the poloidal magnetic field lines in the $r$-$z$ plane superimposed on the background toroidal magnetic field $B_{\phi}$ (left-hand panel), and the ratio of the gas to magnetic pressure $\beta$ (right-hand panel). Regions of high $\beta$ are dominated by gas pressure, while those with low low $\beta$ are supported by magnetic pressure.  The merger and accretion disk themselves remain relatively weakly-magnetized ($\beta \gg 1$), whereas the disk corona and biconical jets are strongly-magnetized ($\beta \la 1$), as previous MRI studies have found \citep {millerstone00}.


\begin{figure*}[th!]
 \label{fig:cartoon}
\begin{center}
\includegraphics[width=0.6 \textwidth]{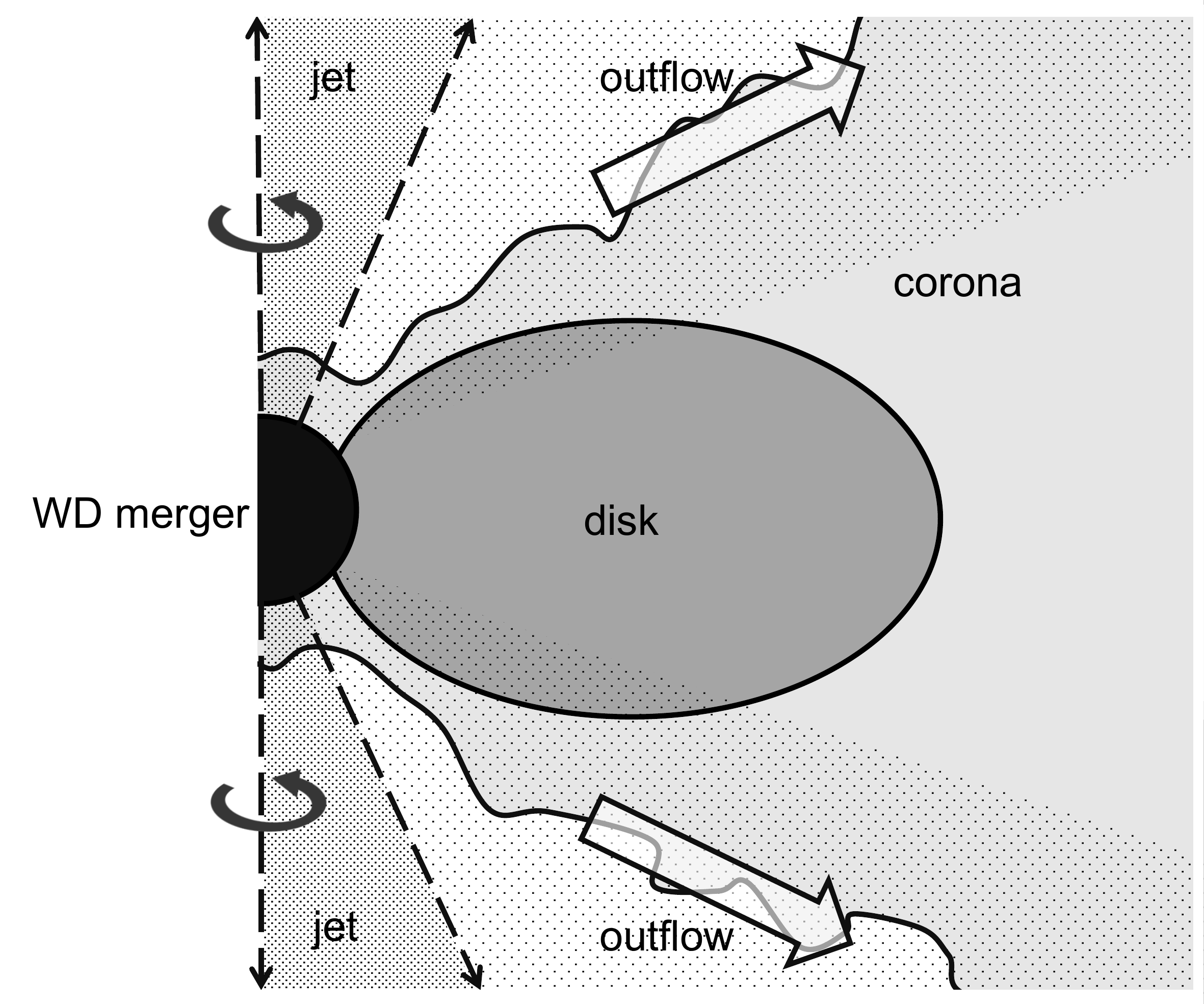}
\caption{A schematic diagram illustrating the key features of our magnetized model, after the development of the MRI. A weakly-magnetized Keplerian disk surrounds the uniformly-rotating white dwarf merger. The development of the MRI within the disk gives rise to strongly-magnetized biconical jets and a magnetized corona. Additionally, strong outflows are driven at the interface of the jet and the corona.}
\end{center}
\end{figure*}

The magnetic field structure in the disk is highly-turbulent and disordered. Loops of low-density, heated magnetic flux rise buoyantly above the accretion disk into the corona \citep {machidaetal00}, where some reconnect through numerical resistivity, thereby heating the coronal region. Some poloidal loops of flux --- which actually  are toroidal in shape in an axisymmetric geometry ---  are long-lived in our simulation, persisting for many local dynamical times. While it is known that these poloidal flux tori are subject to a wide variety of instabilities in 3D, including the kink and interchange instabilities, both the toroidal field and the differential shear in the disk \citep {spruitetal95} may help stabilize these even in full 3D.

\begin{figure*}[th!]
\begin{center}
  \subfigure[$E_{\rm mag}$]{
    \label{fig:emag}
    \includegraphics[width=0.49 \textwidth]{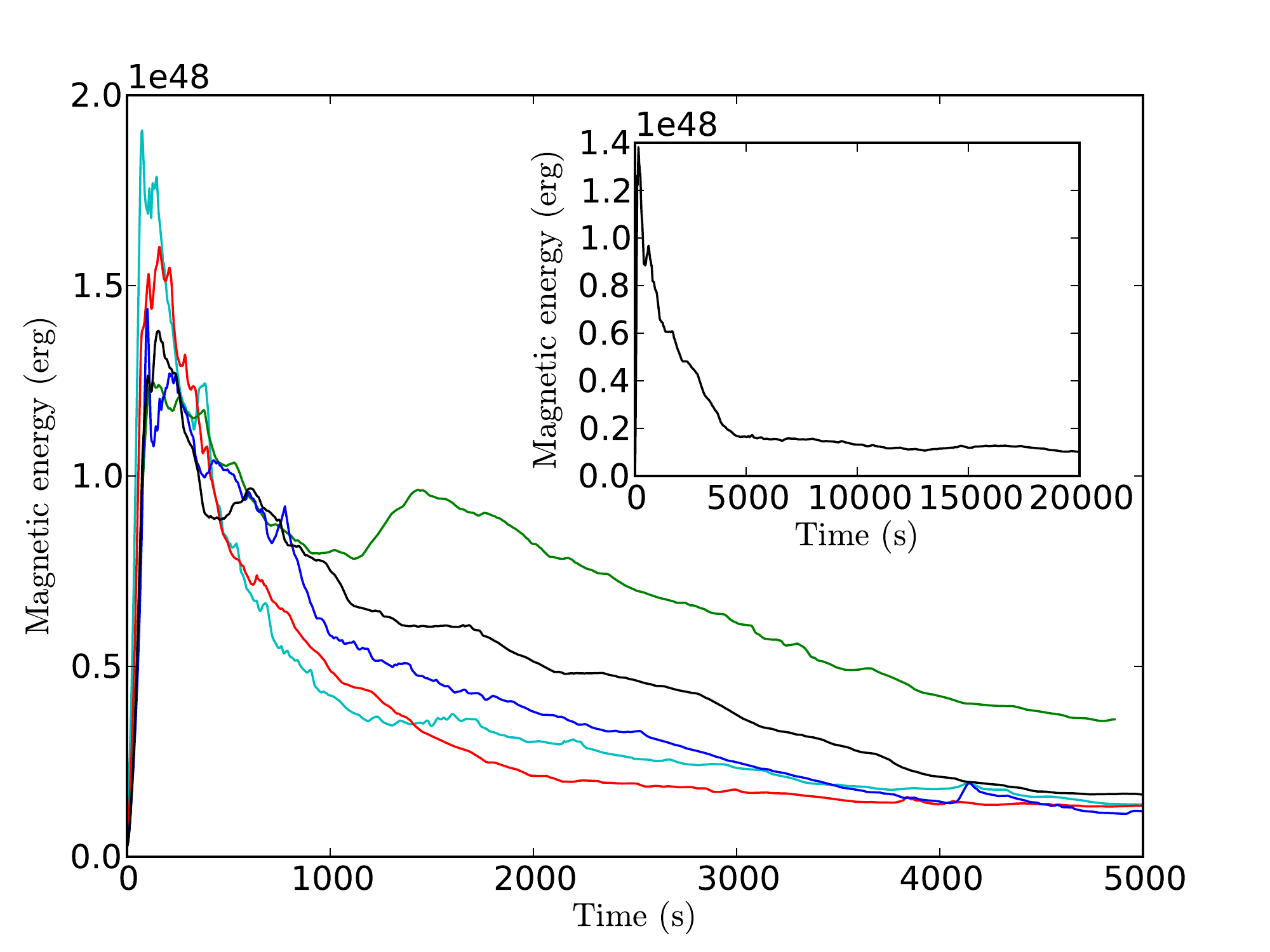}}
  \subfigure[$\langle \alpha_m \rangle$]{
    \label{fig:alpham}
    \includegraphics[width=0.49 \textwidth]{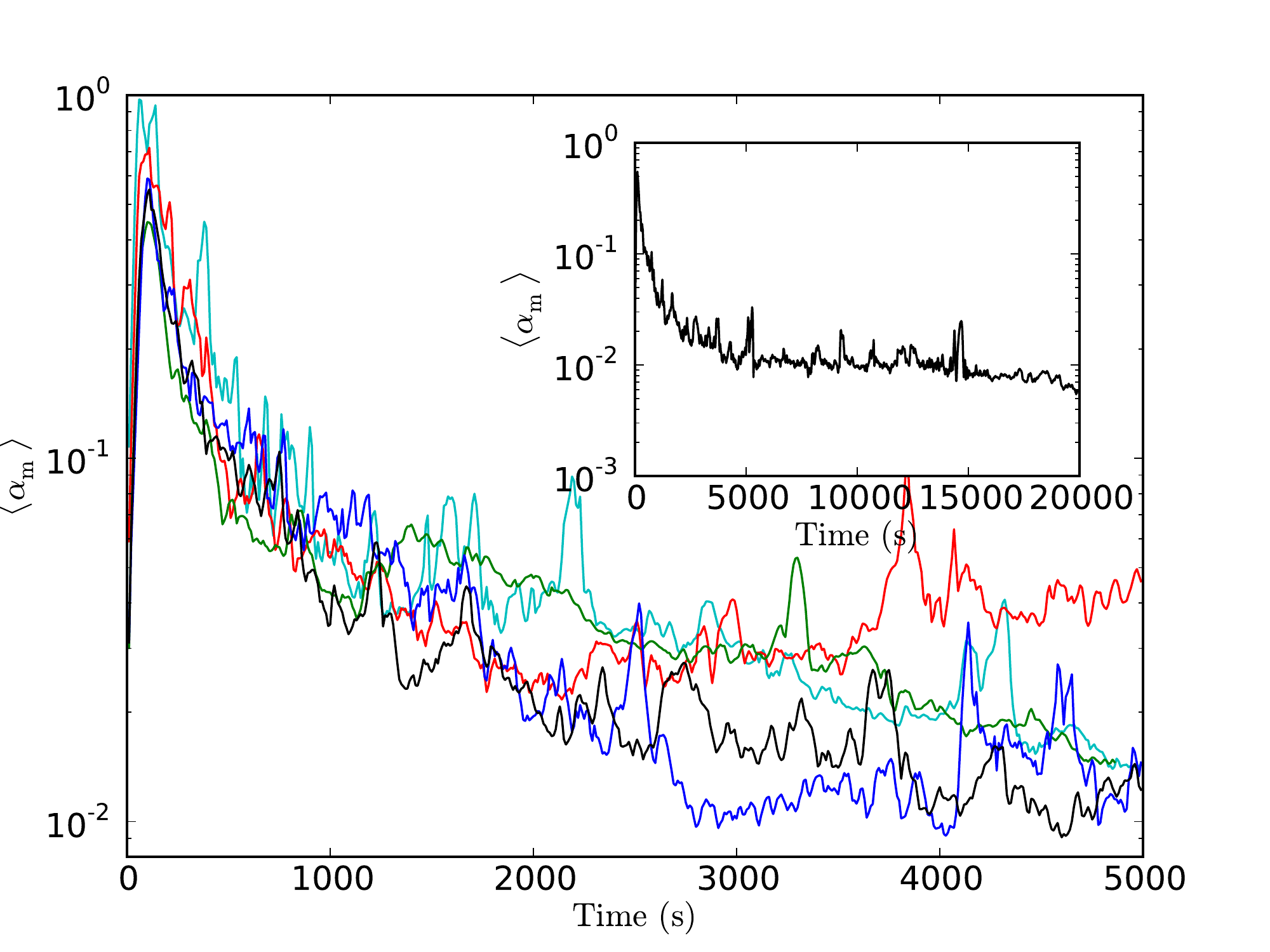}}
\caption{a) The evolution of the global magnetic energy for all models (S-bh black, H-bh blue, L-bh green, S-bl cyan, S-bm red). The left panel inset shows the evolution of the magnetic energy for our standard model over the entire duration of the simulation. b) The effective Shakura-Sunyaev magnetic alpha coefficient $\langle \alpha_m \rangle$ (see text for definition) for all runs, with the same color notation as the panel a. The right panel inset shows the evolution of $\langle \alpha_m \rangle$ over the entire simulation for our standard model.}
\label{fig:emagalpham}
\end{center}
\end{figure*}

In contrast, biconical axial outflows carry open field lines away from the merger. The biconical region is strongly magnetized and heated to $T \sim 10^8$ K, as is clearly seen in figures~\ref{fig:logtemp} and \ref {fig:logbeta}. A strong outflow is driven at the interface of this region with the magnetized corona, similar to previous MRI studies of black hole accretion disks \citep {devilliersetal05}. However, we find that this interface region, which  is Kelvin-Helmholtz unstable,  varies significantly in location and shape over the duration of the simulation. Moreover, although there is a net outflux of both mass and magnetic flux from the simulation domain, there are also thin returning flows, driven by reconnection, which infall onto the disk and merger, similar to that seen in previous work \citep {igumenschevetal03}.

Throughout this paper,  we will have the need to separate the domain into the white dwarf merger, rotationally-supported disk, and magnetized corona in order to determine the properties of each of these regions individually --- see figure~2.
 We characterize each of these regions based upon a physical criterion: specifically, we define our magnetized coronal region to be dominated by magnetic pressure ($\beta < 1$), with the remainder of the weakly-magnetized gas $(\beta \ge 1$) divided into both the white dwarf merger and the disk.  The white dwarf merger is defined to consist of the region primarily supported by pressure, and not rotation ($P  > {1 \over 2} \rho v_{\phi}^2$), while the disk  is rotationally-supported (${1 \over 2} \rho v_{\phi}^2 \ge P$). Furthermore, our definition does not separate out the biconical jets and outflows into distinct additional components; the coronal region includes both the biconical jets and outflows.

\subsubsection{Growth of the Magnetic Field and Magnetic Stress}
\label {sec:fieldgrowth}

We expect the disk to be strongly unstable to the magnetorotational instability, and indeed, we confirm this to be the case. In figure~\ref {fig:emag}, we show the development of the magnetic energy, for each run. For higher resolutions, we achieve a higher peak magnetic energy,  though the runs exhibit a trend towards convergence in the peak magnetic energy with increased resolution. Specifically, the difference in the peak magnetic energy of run H-bh ($E_{\rm mag}^{\rm peak} = 1.44 \times 10^{48}$~erg) and S-bh  ($E_{\rm mag}^{\rm peak}  = 1.38 \times 10^{48}$~erg) is 0.46 that of the difference between the next two lowest-resolution models,  S-bh and L-bh ($E_{\rm mag}^{\rm peak}  = 1.25 \times 10^{48}$~erg).


Next, we explore the role of stresses within our model.  The $r$-$\phi$ component of the stress tensor,
\begin {equation}
T_{R \phi} = \rho \delta v_R \delta v_{\phi} - {B_R B_{\phi} \over {4 \pi}},
\end {equation}
governs the transfer of angular momentum in the disk.  The first term of the stress tensor is the Reynolds stress, and the second, the Maxwell stress. Here $\delta v_R$ and $\delta v_{\phi}$ are the fluctuations of the radial and azimuthal velocity components.  Analytically, during the linear growth phase of the MRI in a near-equilibrium disk, these fluctuations are the departures of the local fluid velocity from a circular orbit --- specifically, $\delta v_R = v_R$ and $\delta v_{\phi} = v_{\phi} - R \Omega (R)$, where $\Omega (R)$ is the disk angular velocity at radius $R$.  However, it is well-known that due to the large turbulent fluctuations in fully-developed MRI turbulence, defining the mean disk velocity is problematic, even in a time-averaged sense, and consequently  there is no unique prescription for specifying the Reynolds stress in a global disk simulation --- see for instance, \cite {hawleykrolik01}.  Here, we focus upon the Maxwell stress as a proxy for the total stress, since 3D simulations of the MRI typically find the Maxwell stress dominates the Reynolds stress by factors of 3--6  --- see, for instance, \citet {davisetal10}. In figure \ref {fig:alpham}, we plot the ratio of the spatial average of the Maxwell stress to the gas pressure, which defines an effective magnetic Shakura-Sunyaev  parameter $\langle \alpha_m \rangle =  \left \langle - \frac{1}{4 \pi}  B_R B_\phi  \right \rangle / \left \langle P_{\rm gas} \right \rangle$, as a function of time, for each run. Here, brackets indicate spatially-averaged quantities over the disk and coronal regions, excluding the white dwarf merger itself.
 
 We find overall good agreement in the Maxwell stresses in all models computed, with all models apart from S-bm converging towards a value of $\langle \alpha_m \rangle \sim 0.01$. Moreover, our standard model, which has been evolved for the longest runtime, has a relatively steady $\langle \alpha_m \rangle \sim 0.01$, indicative of sustained accretion and angular momentum transport. 
 
The magnetized stress reflects the turbulent dynamics of the MRI in the disk, and like many properties of turbulent systems, is best understood in terms of a stochastic behavior with large departures from mean values. To better quantify this behavior, we have computed the time- and spatially-averaged $\alpha_m$, which we define as 

\begin {equation}
\langle \langle \alpha_m \rangle \rangle  = {1 \over T} \int_{t_0}^{t_0 + T} \langle \alpha_m (t) \rangle dt
\end {equation}
Here double angle-brackets indicate both spatial and time-averaging over an interval of $T$, beginning with $t = t_0$. We time-average the late-time evolution of figure \ref {fig:alpham},  taking $t_0 = 1000$ s and $T = 5000$ s. We find a trend towards convergence  in $\langle \langle \alpha_m \rangle \rangle$ for our high $\beta$ model with increased resolution -- with values ranging from 0.033 (L-bh) through 0.020 (S-bh) to .023 (H-bh), but with some sensitivity at fixed resolution to the initial choice of $\beta$. In particular, we find $\langle \langle \alpha_m \rangle \rangle = 0.058$ and $0.027$ for $\beta = 1000$ and $\beta = 500$, respectively.

\begin{figure*}[th!]
\begin{center}
  \subfigure[Mean White Dwarf Merger Magnetic Field]{
    \label{fig:meanmagnetic}
    \includegraphics[width=0.49 \textwidth]{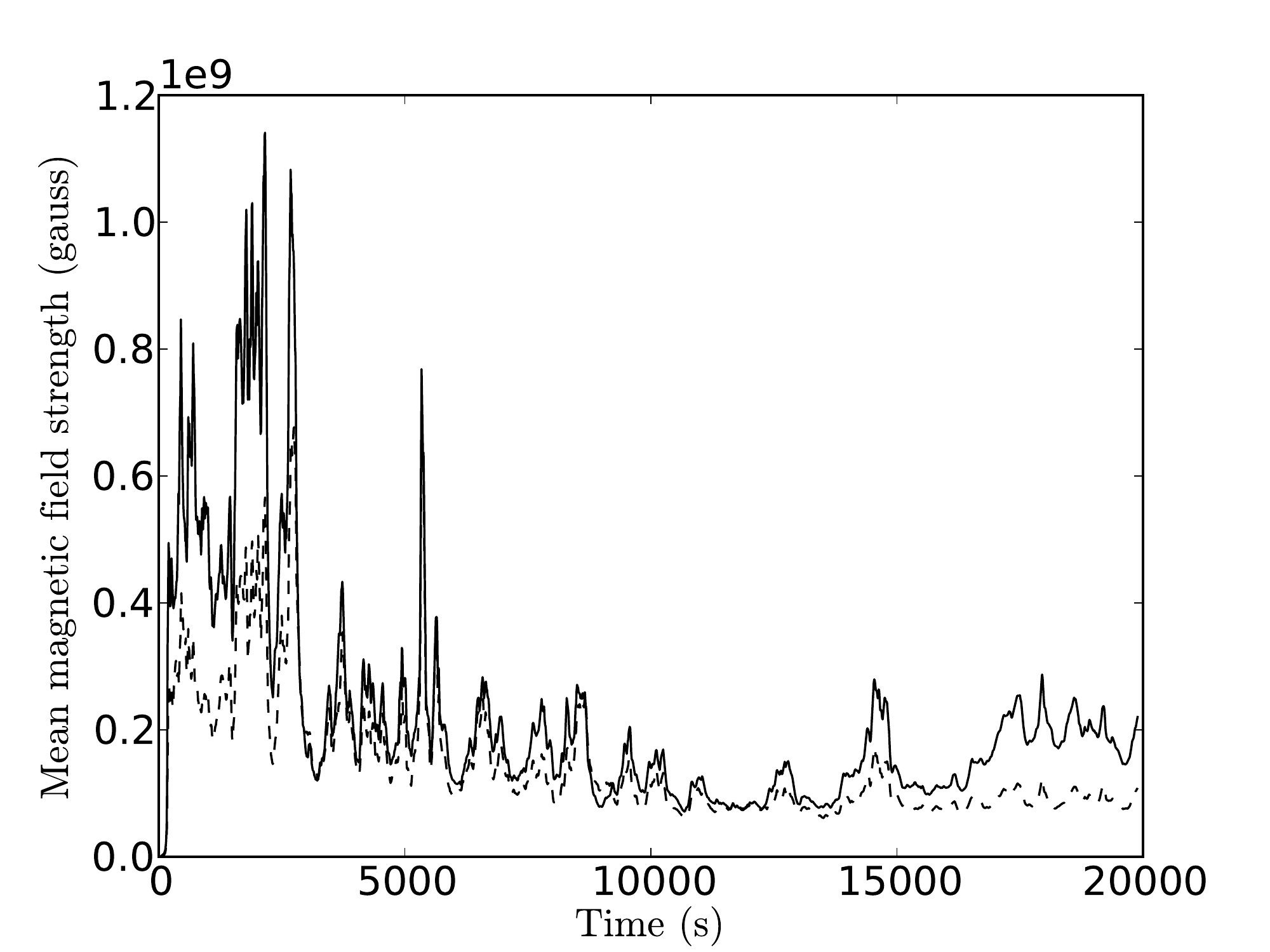}}
  \subfigure[Ratio of White Dwarf Merger Mean $B_t / B_p$]{
    \label{fig:ratiomagnetic}
    \includegraphics[width=0.49 \textwidth]{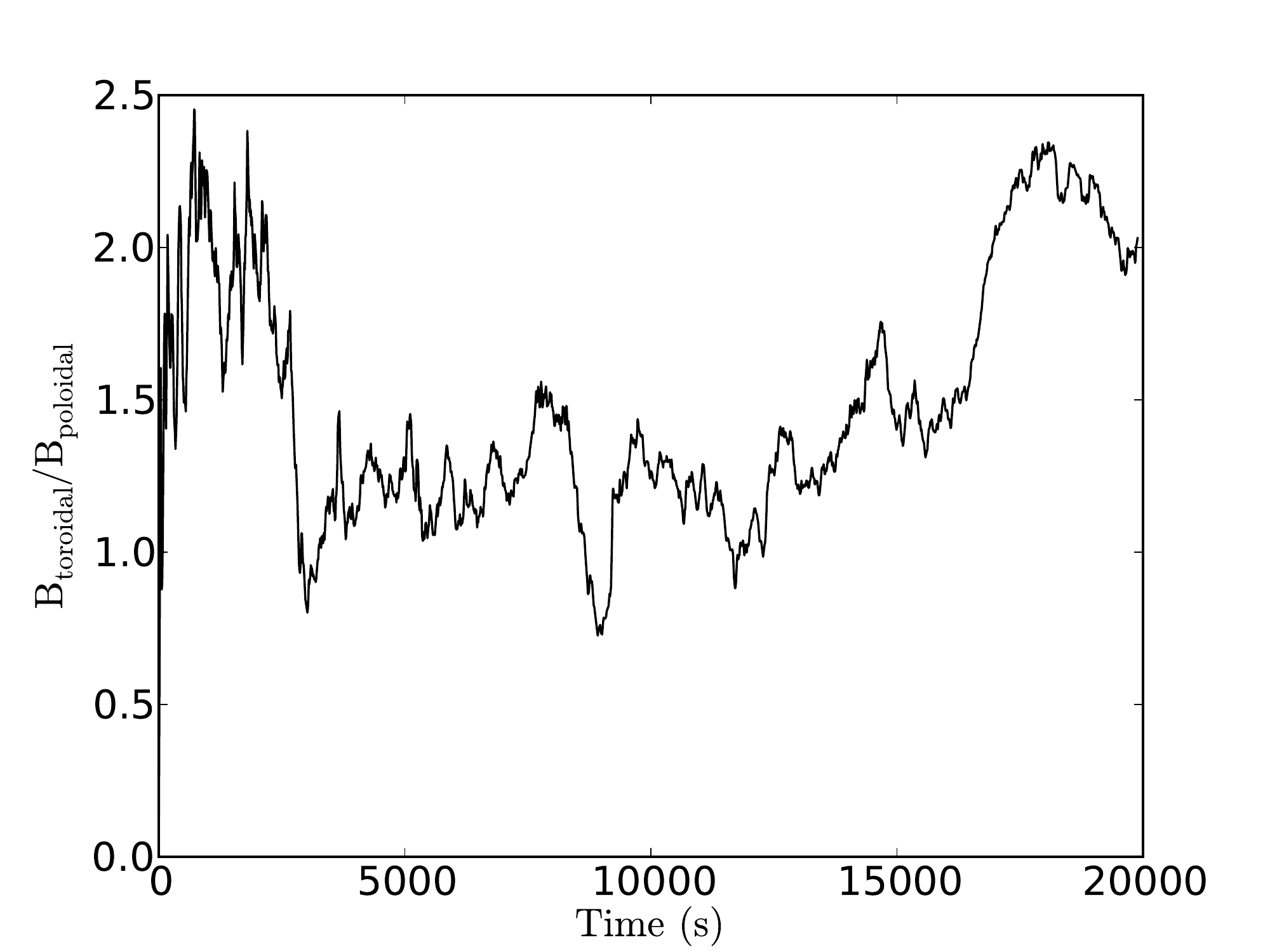}}
\caption {a) The evolution of the mean magnetic field within the white dwarf merger versus time, shown for both the mean toroidal (solid) and mean poloidal (dashed) lines for model S-bh. b) The ratio of mean toroidal to mean poloidal magnetic field within the merger, versus time, again for model S-bh.}
\label{fig:mergerfield}
\end{center}
\end{figure*}

 The white dwarf merger, which is initially weakly magnetized with a mean, purely poloidal magnetic field of $2.8 \times 10^5$~G, becomes rapidly magnetized over several inner rotational periods, as the MRI develops in the disk and magnetic flux is advected into the merger --- see figure~\ref {fig:meanmagnetic}. The field strength is time-variable, particularly at early times, which is expected given the turbulent non-steady nature of the MRI in the disk ---  just as the mass accretion is highly variable, so too must be the advection of magnetic flux into the merger. At late times, the total mean magnetic field strength within the merger is $\sim 2 \times 10^8$~G, typical of HMFWDs.   The ratio of the mean toroidal field strength to the mean poloidal field strength is shown in figure~\ref {fig:ratiomagnetic}. The final field is predominantly toroidal, with a mean value of $B_t / B_p \sim 1.5$.  Both the field strength and the geometry, as traced by the toroidal to poloidal field strength ratio, are highly time-variable, indicative of a disordered interior magnetic field. Given our limited resolution within the white dwarf merger of roughly 60 cells in our standard model S-bh, our final field strengths are very likely limited by numerical resistivity. Our results do, however, demonstrate that HMFWD field strengths may be achieved through the double-degenerate channel.

The global magnetic energy $E_{\rm mag}$ exhibits a slow decay at late times. About $7 \times 10^{47}$~erg, or roughly half of the total drop in magnetic energy is actually due to the outflow of magnetic flux from the problem domain in our model S-bh. At late times, more magnetic flux is outfluxed than the net magnetic energy generated within the domain by the MRI, as the MRI saturates and slowly winds down, and turbulent magnetic energy is dissipated through reconnection. The net result is a net decrease in magnetic field energy on the problem domain. 

We note that while the simulations presented here correspond to a relatively special case of an equal-mass white dwarf merger, which results in a central peak temperature. In the more general case of an unequal mass merger, the peak temperature will occur in the nearly spherically-symmetric, hot, convective region surrounding the primary white dwarf, in which the MRI is also expected to grow rapidly \citep {garciaberroetal12}. However, the key point is that the MRI is expected to be the generic outcome of a merger, and the calculations presented here demonstrate this concretely for this specific model.

Moreover, we can estimate the lifetime of the magnetic field, assuming that there is no further dynamo action present, and that the magnetized, accreted disk material is spread into a spherical shell surrounding the white dwarf \citep {nordhausetal11}. In this case, the timescale against Ohmic decay via Spitzer resistivity is $\sim$ 200 Myr for a $10^8$ K, $0.2\ M_{\odot}$ disk. Thus we expect the surface fields to be strong enough to be visible as a HMFWD for some time beyond merger. Consequently, mergers may account for the newly-discovered class of hot DQ white dwarfs, of which roughly 70\% are strongly-magnetized, a level far exceeding that of the field \citep {dunlapetal10, dufouretal11, lawrieetal13, williamsetal13}.

\subsubsection{Mass Accretion and Mass Outflow}
\label {outflow}

The stresses developed by the MRI drive mass accretion through the disk. Additionally, previous studies have demonstrated that the turbulence produced within the disk can develop  coronal mass outflows \citep {machidaetal00, millerstone00, suzukiinutsuka09, flocketal11}.  The implications of significant mass outflow are particularly profound for early-time optical, X-ray, and radio observations of SNe Ia, which are sensitive probes of the circumstellar environments surrounding the progenitor systems. 

A key topic of observational interest is the recent discovery of narrow ($\sim 10$ km/s) Na I doublet absorption lines in the light curves of some normal SNe Ia, originating from the circumstellar medium (CSM) surrounding the SNe Ia progenitor \citep {patatetal07, foleyetal12}.  Originally, these lines were interpreted as likely supporting a symbiotic progenitor channel for the SNe Ia, with the implicit assumption that other progenitor channels were likely to have a very sparse or nonexistent CSM. However, recent work by several groups have questioned this assumption by demonstrating that a CSM with properties similar to the observed narrow NaID lines can be formed during the late snowplow phase of radiative, fast outflows from a variety of other models. The surface escape speed of a white dwarf is typically $\sim$ several $\times 10^3$ km/s, which  is much greater than CSM velocities implied by the narrow NaID. However, {\it provided} that the SNe event is significantly delayed after the emergence of the outflows -- by hundreds to tens of thousands of years depending on the initial outflow speed and the assumed density of the interstellar medium surrounding the progenitor --  a number of various progenitor channels may also give rise to a CSM with properties quite similar to the observed NaID lines, through the basic physics of the momentum-conserving snowplow phase of the outflow.  For instance, \citet {shenetal13} demonstrate that multiple shells can be formed during the mass transfer between a He+C/O double white dwarf binary preceding a sub-Chandrasekhar double-detonation event, and  \citet {sokeretal13} demonstrate that multiple shells may also arise in a core-degenerate scenario of the merger of a white dwarf with the core of an asymptotic giant branch phase star. Additionally, subsequent to the initial submission of this paper, \citet {raskinkasen13} have found that a tidal tail of $\sim$ several $\times 10^{-3} M_{\odot}$ of mass is unbound during the final white dwarf binary merger itself, very similar to the amount in the original SPH simulations upon which our results here are based  \citep {lorenaguilaretal09}.

We characterize the initial mass of the white dwarf merger, disk, and magnetized corona by analyzing each of these at an early point in the simulation ($t < 500$~s), where the MRI is fully-developed, but prior to significant post-merger mass changes. At these early times, the white dwarf merger mass is $0.96\, M_{\sun}$ by our specified criteria; the disk and the corona account for $0.20\, M_{\sun}$ and $0.04\, M_{\sun}$, respectively. 

In addition to the mass accretion from the disk onto the white dwarf merger, a significant fraction of the total mass of the disk is lost, primarily through accretion as well as an outflow driven near the interface of the corona with the biconical jet with a small fraction exceeding the escape speed.
Over the duration of our standard model, we determine that the disk has lost nearly 90\% of its initial mass over the duration of the simulation, through a combination of accretion and outflow. Of this amount, just over 82\%, or $0.16\, M_{\sun}$, is accreted onto the white dwarf merger, with the remainder either transferred into the corona or outfluxed and lost from the domain. In total, nearly $0.06\ M_{\sun}$ is outfluxed from the domain and subsequently lost to the simulation. However, the vast majority of this mass remains gravitationally-bound in our simulation (see table \ref{tab:sims}), and is likely to be re-accreted. Only about $10^{-3} M_{\odot}$ of the outflow is gravitationally-unbound over a $2 \times 10^4$ s duration,  and will be completely lost from the system. The mean ejection velocity is $2600$ km/s, which is much in excess of the escape speed at the top and bottom edges of the domain of $\sim 1600$ km/s.

In order to quantify the angular distribution of the mass outflow, we have computed the angle of the outflow of each unbound cell, relative to the vertical axis. While matter is ejected in both the radial and vertical directions, the magnetically-driven outflow is preferentially-driven along a 50$^{\circ}$ angle relative to the vertical direction, which is the mean of the mass density outflux angular distribution -- see figure \ref {fig:OutflowAngle}.  We note that the total mass outfluxed, including both bound and unbound mass, from the domain over the duration of the simulation varies by a factor of roughly 2, from $0.027\ M_{\odot}$ for model S-bl to $0.058\ M_{\odot}$ for model S-bm, as shown in table~\ref {tab:sims}.

\begin{figure*}[th!]
\begin{center}
\includegraphics[width=0.8 \textwidth]{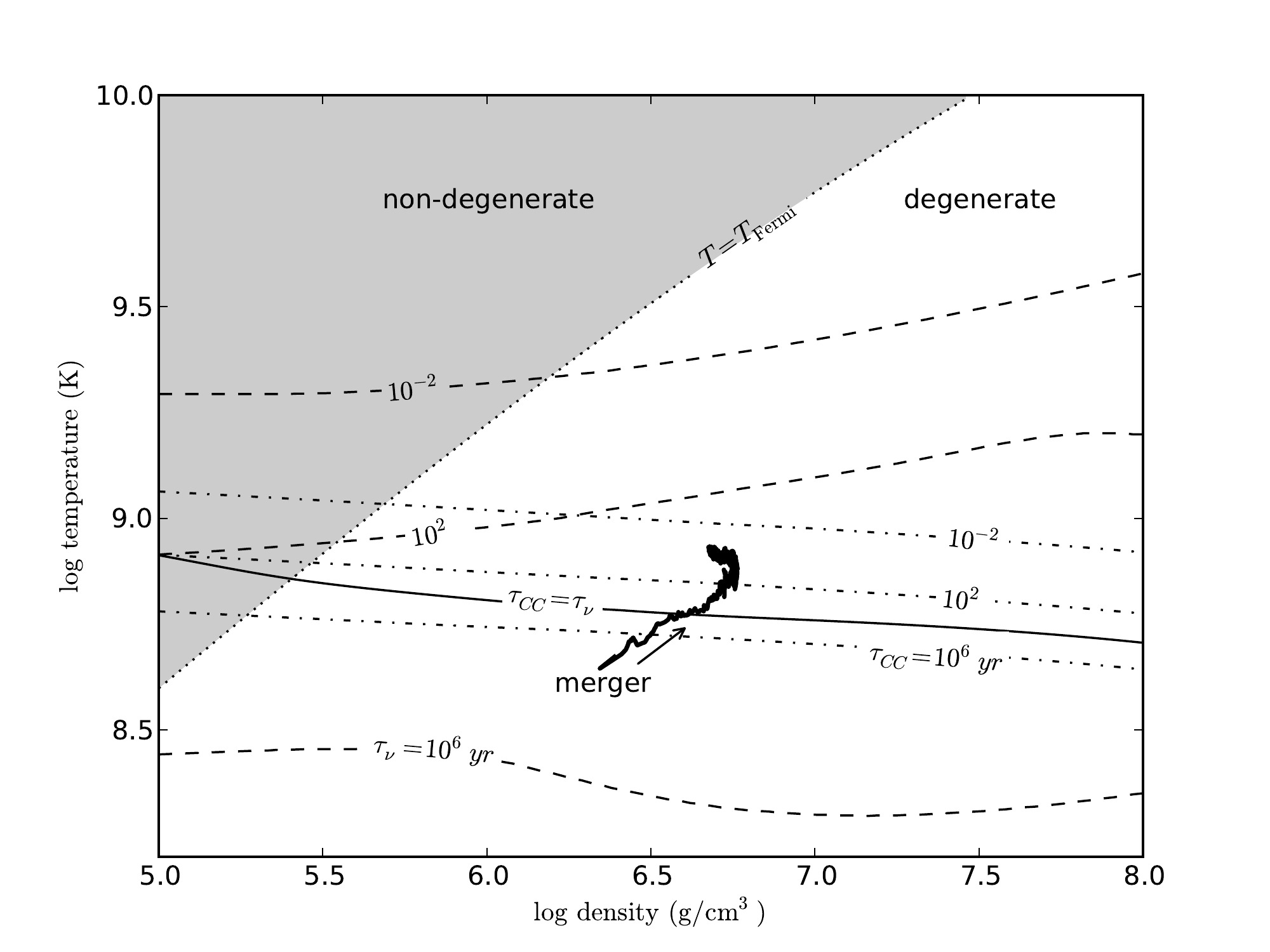}
\caption{A phase plane of the central conditions of the simulated white dwarf merger for model S-bh. See text for description.}
 \label{fig:logtemp_vs_logdens}
\end{center}
\end{figure*}

The total momentum of ejected material in the tidal tail is quite similar to the momentum ejected in our wide, magnetically-driven outflow, though the tidal tail material is preferentially ejected in the plane of the merger with a slightly lower characteristic velocity. For a significant delay of $\sim 10^4$ yr between the merger and  a possible SNe event, \citet {raskinkasen13} demonstrate that these tidally-driven outflows may produce distinct observational signatures in narrow NaID lines.  For such a delay subsequent to the initial merger, we expect that magnetically-driven winds will act in concert with tidal tails, with both the tidal tails and the magnetically-driven outflows sweeping up the interstellar medium to produce multiple shocked, asymmetric  shells in narrow NaID lines. These predicted shells are perhaps similar to existing observations of asymmetric NaID shells \citep {forsteretal12}, or of multiple NaID shells in the PTF11kx system \citep {dildayetal12}, depending on the viewing angle between the SNe event and the observer.




\subsubsection{Spin-Down of White Dwarf Merger}

A further key question is the possible influence of the magnetic field upon the spin of the white dwarf merger. Previous studies of core-collapse supernovae have suggested that the outward transport of angular momentum results in a  spin-down of the protoneutron star and its surroundings --- see for instance, \citet {thompsonetal05}. In the context of the double-degenerate model of SNe~Ia, a loss of rotational support of the central white dwarf merger may yield additional or perhaps even dominant compression beyond that provided by  accretion from the  disk.

\begin{figure*}[th!]
\begin{center}
  \subfigure[$\Omega$ vs. $r$]{
    \label{fig:omegavsr}
\includegraphics[width=0.49 \textwidth]{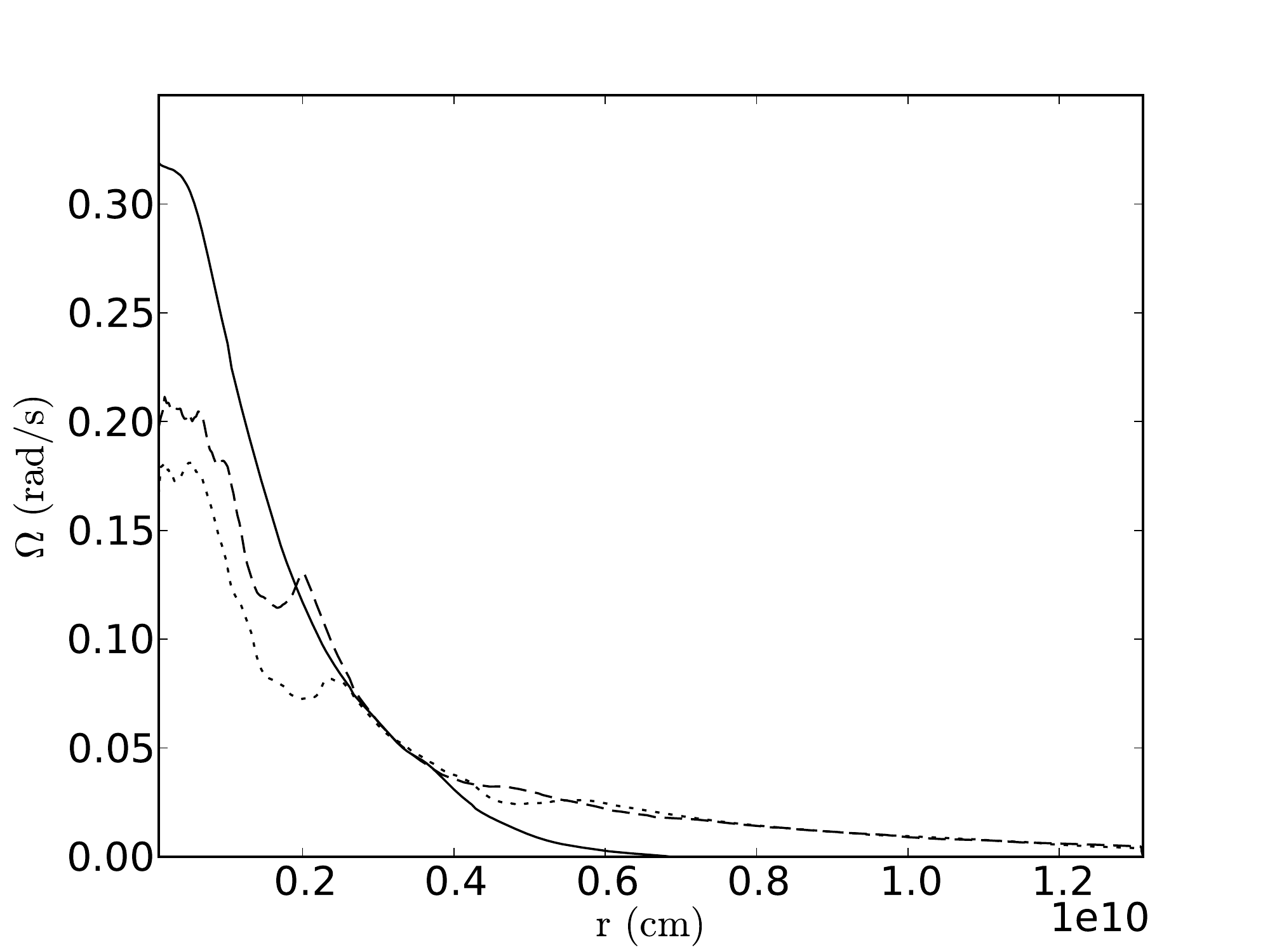}}
  \subfigure[$J$, $\Delta J$ vs. time]{
    \label{fig:braking}
    \includegraphics[width=0.49 \textwidth]{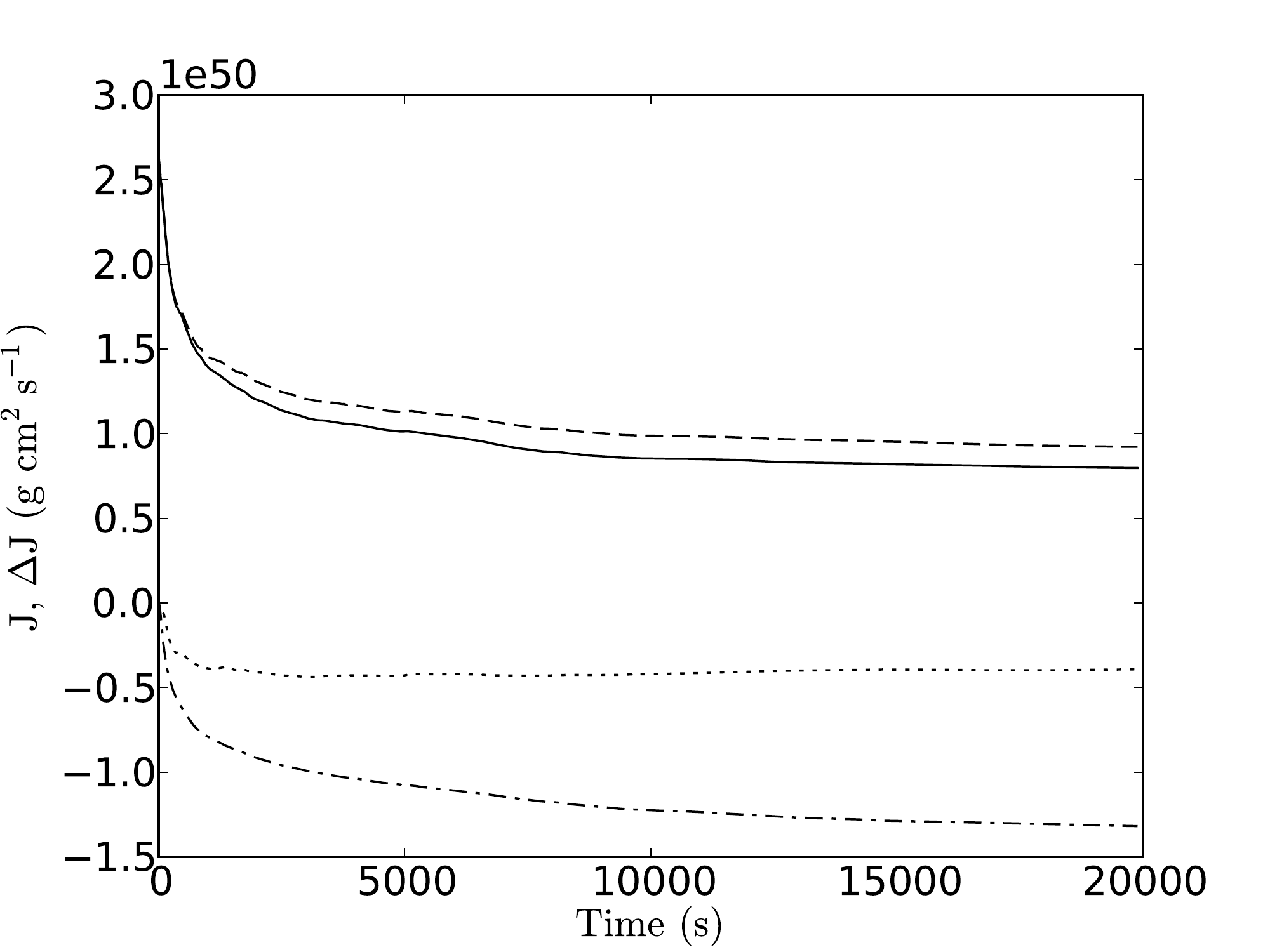}}
\caption{a) A plot of the angular velocity $\Omega$ versus cylindrical radius $r$, shown at various times --- solid curve ($t = 0$~s), dashed curve ($t$ = $10^4$~s), and dotted curve ($t = 2 \times 10^4$~s) for model S-bh.  b) A plot of the angular momentum and computed changes in angular momentum of the white dwarf merger vs. time for model S-bh. The solid line represents the actual simulated spin, whereas the dashed line is the computed spin based upon the combined effects of the Maxwell and Reynolds stresses, whose cumulative effect is shown in the dash-dotted and dotted curves, respectively.}
\end{center}
\end{figure*}

\begin{figure*}[th!]
\begin{center}
\includegraphics[width=0.8 \textwidth]{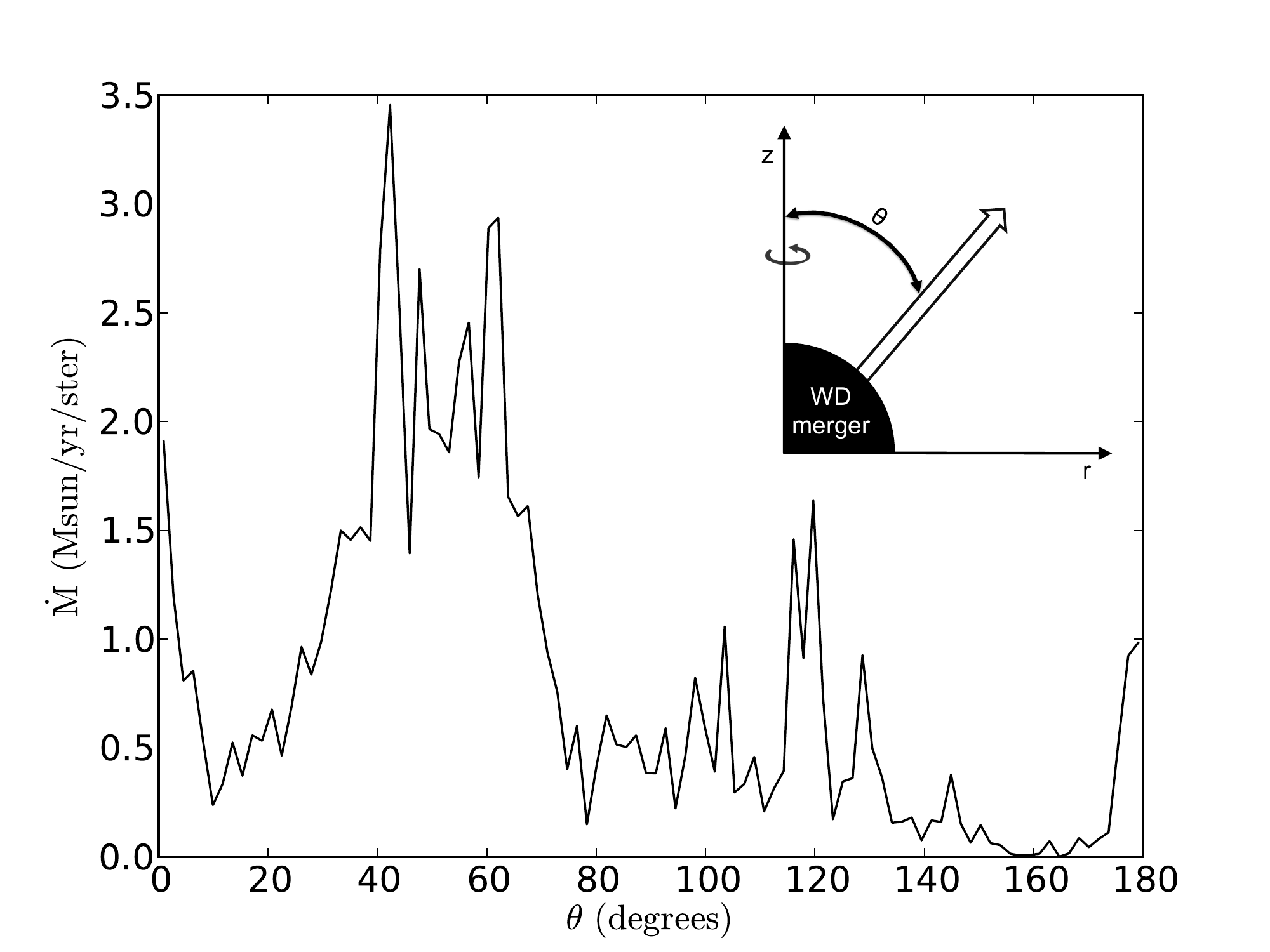}
\caption{The distribution of the outflow angle for matter ejected by the magnetized outflow, in units of $M_{\odot}$  yr$^{-1}$ ster$^{-1}$,  plotted versus spherical angle $\theta$. The inset shows the geometry of the outflow angle.} 
 \label{fig:OutflowAngle}
\end{center}
\end{figure*}

We find the central white dwarf merger is spun down significantly on a very rapid timescale, due primarily to the development of Maxwell stresses at the boundary of the white dwarf merger.  In particular, we find that the merger, whose initial angular momentum is $2.6 \times 10^{50}$~g~cm$^3$~s$^{-1}$, spins down significantly to $8.0 \times 10^{49}$~g~cm$^3$~s$^{-1}$ by the end of the standard model simulation. This amounts to a braking timescale of  $J /  | \dot {J} | \simeq$ several $\times 10^4$~s. In figure~\ref {fig:omegavsr}, we plot the vertically-averaged angular velocity $\Omega$ as a function of cylindrical radius $r$. The white dwarf merger is seen to be in solid-body rotation at $r < 10^9$~cm, with the outer portion of the domain is in Keplerian rotation. We note the tapering of the angular velocity plot at radii $r > 6 \times 10^9$~cm at $t = 0$~s is an artifact of the initial SPH particle distribution, which lacked any particles in this region. Additionally we note that the shear in the innermost portion of the disk, near $r = 2 \times 10^9$~cm, flattens out as the free energy available in the shear is tapped to drive the MRI.  As time evolves, the merger is seen to spin-down significantly, from $\Omega \sim 0.3$~s$^{-1}$ initially to almost half of that value, $\Omega \sim 0.17$~s$^{-1}$, finally, at the end of the run. 
 
In order to verify that the simulated spin-down of the merger is indeed physically accurate, and not the consequence of numerical errors, we have computed the spin-down expected from the conservation of angular momentum and compared this to the measured spin-down. In particular, in axisymmetry, the inviscid angular momentum evolution equation can be written as 

\begin {equation}
{\partial \over \partial t} (\rho R v_{\phi} ) + {\bm \nabla} \cdot R \left[\rho v_{\phi} {\bm v} - {B_{\phi} \over 4 \pi} {\bm B_p}     \right] = 0
\end {equation}
%
Here, ${\bm B}_p$ is the poloidal magnetic field. The second term represents the divergence of the angular momentum flux, 
\begin {equation}
{\bm F}_{\phi} = R \left[\rho v_{\phi} {\bm v} - {B_{\phi} \over 4 \pi} {\bm B_p}     \right] 
\end {equation}
In post-processing, we integrate the divergence of the angular momentum flux ${\bm F}_{\phi}$ inside the merger, and compare it with the actual angular momentum evolution of the merger as obtained by the full simulation. For the purposes of this computation we define the merger  as the fixed region with a distance less than $1.5 \times 10^4$~km from the domain center, which is consistent with our previous definition based upon pressure support. The result is shown in figure~\ref {fig:braking}. The cumulative effect of the magnetic and hydrodynamic stresses are integrated over volume and time to produce a net change in angular momentum, shown in the dash-dotted curve. The Maxwell stresses are roughly three times more significant than the Reynolds stress, similar to previous MRI studies, e.g.,  \citet {flocketal11} and \citet {davisetal10}. 

The sum of these predicted net changes in angular momenta are plotted, and added to the initial angular momentum, as the dashed curve, which compares with the actual simulated angular momentum, as shown in the solid curve. The results agree to within 10\%, which establishes that the spin-down is physically driven by MRI-generated Maxwell stresses, and is not the result of numerical or artificial viscosity. Furthermore, by subsampling the output interval, we have determined the dominant error in the calculation to be the finite output cadence of 10~s. Thus, while our calculation places the upper-bound of the numerical errors at $\sim 10\%$, the true numerical error is very likely to be much smaller.



\subsubsection {Nuclear Burning}
\label {sec:nuclearburning}

The question of whether the merger will initiate a detonation, and possibly result in a SNe~Ia, hinges crucially upon the peak nuclear burning rate. Because the equal-mass merger case produces a central peak in temperature, the burning rate achieves its highest value at the merger center. In figure~\ref {fig:logtemp_vs_logdens}, we plot the central conditions for the white dwarf merger in the $\log T - \log \rho$ plane, as a thick solid line.  Also shown are equicontours of both the carbon-burning (dot-dashed) and neutrino timescales (dashed), at values of $10^6$~yr, $10^2$~yr, and $10^{-2}$~yr. The critical condition for thermal runaway, at which the timescale for carbon burning is equal to that of neutrino cooling, is shown as a thin solid line.  The degeneracy boundary in the temperature-density plane is demarcated by the dotted where the temperature equals the Fermi temperature: $T = T_{\rm F}$.  The central conditions of the merger remain partially, though not fully, degenerate throughout the evolution.   At the beginning of the simulation, at $t = 0$~s immediately after the initial merger, the neutrino cooling time is shorter than the carbon-burning time at the center of the merger. As the white dwarf spins down and accretes mass from the disk, the central region compresses further, and the central conditions become thermally unstable, with the timescale for carbon burning shorter than neutrino cooling at around $t_{\rm CC} = 10^4$~yr. At this point, the central region of the white dwarf has ignited.

After ignition, the central temperature of the merger continues to rise, reaching a peak temperature $T \sim 10^9$~K. The burning timescale at this temperature is roughly 1~yr, which is much longer that what we can feasibly follow in this set of simulations. Additionally, the maximum temperature is not converged, and shows a systematic increase of roughly 20\% both as the resolution increases, or as $\beta$ decreases. This behavior is consistent with the maximum temperature for a sharply-peaked temperature profile,  which becomes coarsened at lower resolution. Further, since decreasing the initial $\beta$ parameter is equivalent to an increase in the effective resolution of the MRI critical wavelength $\lambda_c$, both trends may be understood on the basis of the effective resolution of the simulation. We return to this issue below in the conclusions.

\section{Comparison With Previous Results}
\label{sec:previousresults}

While there is a large body of work on the final stages of the merger of white dwarf binaries \citep {mochkovitchlivio89, mochkovitchlivio90, rasioshapiro95, segretainetal97, guerreroetal04, dsouzaetal06, yoonetal07, motletal07, pakmoretal10, danetal11, zhuetal11}  there are relatively few in-depth studies of the post-merger phase of evolution.  Several classic spherical studies of \citet {saionomoto85}, \citet {saionomoto98}, and \citet {saionomoto04}, follow the evolution of a spherical remnant accreting near the Eddington limit, and concluded that the off-centered ignition would lead to an accretion-induced collapse. Later, \citet {yoonetal07} followed six white dwarf mergers using SPH, varying both the mass ratio and the number of particles. Their most-advanced model consisted of a super-Chandrasekhar SPH simulation of $0.6\, M_{\sun} + 0.9\, M_{\sun}$ white dwarf binary, which  was followed through merger and a short time (300~s) after merger using inviscid hydrodynamics.  Additionally, \citet {shenetal12} and \citet {schwabetal12} follow the post-merger evolution of a range of super-Chandrasekhar mass white dwarf binaries in 1D spherical and 2D spherical geometry, respectively, using a Shakura-Sunyaev viscosity prescription for the transport of angular momentum. The comparison of our work to these previous models is complicated by our focus upon an equal-mass sub-Chandrasekhar case, whereas most prior models in these studies focus upon unequal mass, super-Chandrasekhar models. We are, however, able to compare the broadest features and underlying physics of all models.

 The morphology of the mergers studied by both \citet {yoonetal07} and \citet {schwabetal12}, which consist of a rotating white dwarf core, a hot envelope, and an accretion disk is in broad agreement with the structure of our initial SPH conditions, with the crucial difference that their unequal mass models have temperatures which peak off-center, as other studies have also shown --- e.g.	\citet {lorenaguilaretal09} and \citet {zhuetal11}. We do note that our initial central peak temperatures are somewhat higher than those found by \citet {zhuetal11} for similar sub-Chandrasekhar mergers, and that this is likely due to differences in the SPH modeling. Additionally, \citet {zhuetal11} find that peak central temperatures are achieved only for non-synchronously rotating white dwarfs, as ours are. 
 

\citet {yoonetal07} parameterize the  end state of their SPH models by mass, and  advance a range of models in a 1D hydrodynamic code incorporating a description of angular momentum transport via hydrodynamic instabilities with an imposed timescale of $10^4 - 10^5$~yr. Our models, which include self-consistent angular momentum transport through the magnetorotational instability, as well as the models of \citet {schwabetal12} using an $\alpha$ model, find that the timescale for accretion of the disk is consistent with the turbulent viscous accretion timescale, and many orders of magnitude shorter than model assumptions of \citet {yoonetal07}. Our findings are, however, consistent with the timescale of turbulent disk accretion suggested by \citet {vankerkwijketal10}. Our findings are also consistent with recent experiments which find that hydrodynamic turbulence is inefficient at transporting angular momentum \citep {jietal06}.  They are generally inconsistent with accretion at the Eddington rate \citep {saionomoto85, saionomoto98, saionomoto04}.


Furthermore, \citet {shenetal12} and \citet {schwabetal12} find that the influence of a spatially-dependent turbulent viscosity ($\alpha \sim 0.03$) drives their mergers to spin down completely over a timescale of $\sim 3 \times 10^4$~s, quite similar to our finding with the full MRI that the white dwarf merger is magnetically braked on a comparable timescale. Moreover, they further find that as turbulent viscosity dissipates rotational energy into heat,  a thermally-supported outer region develops, also broadly similar to our heated corona. Additionally, \citet {schwabetal12} find evidence for  sustained carbon  burning, as we do in our models. 

However, there is a subtle but important distinction in the physics of the angular momentum transport and heating mechanisms at work here. The Shakura-Sunyaev prescription posits that shear energy is {\it locally} dissipated as heat energy over a viscous timescale. In contrast, in a magnetized system, the conversion of shear energy in the disk does not directly lead  to heating. Instead, under the action of the MRI, shear energy is converted into magnetic energy, which is subsequently buoyantly displaced into the corona. The corona is ultimately heated through the {\it non-local} dissipation of magnetic energy into heat energy. In our models, this is accomplished through numerical resistivity, though in reality the magnetic dissipation will occur through both physical resistivity and rapid reconnection. Similarly, the inner white dwarf merger is heated under compressional work from both accretion and spin-down. The end result in our case  is a highly intermittent and non-uniform thermal structure --- see figure~\ref {fig:logtemp}. Indeed, about $7 \times 10^{47}$~erg, or roughly half of our peak magnetic energy, is outfluxed entirely from our problem domain.  In contrast, \citet {schwabetal12} find their outer envelope is smoothly heated through the action of turbulent viscosity, which is the direct result of the conversion of local shear energy into heat. Further magnetized models will be able to determine the precise extent to which the thermal structure of the hydrodynamic and magnetic models differ, and identify possible implications for nuclear burning and detonation.

The most significant difference, however, between  \citet {schwabetal12} and the models presented here relate to mass outfluxes. \citet {schwabetal12} find that mass outflows are limited to less than $10^{-5} M_{\sun}$ in their standard $0.6\, M_{\sun} + 0.9\, M_{\sun}$ model. In contrast, our sub-Chandrasekhar model drives a vigorous outflow of unbound mass two orders of magnitude greater.\footnote {However, in an unpublished $0.75\, M_{\sun} + 0.75\, M_{\sun}$ run, they do report a thermally-driven expansion of their white dwarf merger, which would amount to $0.1\ M_{\sun}$ beyond our domain (private communication), which is similar to the total amount of outflux we see in our lower mass model.}  Our results are consistent with global MRI studies, which often find significant outflows --- see for instance, \citet {devilliersetal05} and \citet {flocketal11}.  The outflows generated in magnetized disks  are fundamentally driven by the buoyancy of the magnetized corona and its interaction with the strongly-magnetized biconical jet. The Shakura-Sunyaev prescription utilized by \citet {schwabetal12} captures the basic aspects of mass and angular momentum transport by specifying the $r$-$\phi$ and $\theta$-$\phi$ components of the stress tensor, but lacks both magnetic pressure and tension and the Lorentz force, which play a crucial role in driving magnetized outflows.
 
Moreover, while our models suggest that magnetorotationally-driven outfluxes may influence the immediate circumstellar environment surrounding the white dwarf merger,  the precise fate of this outfluxed matter is underdetermined in our simulations, which do not capture the sonic surface ($R_S = 2 G M / c_s^2 \simeq 10^{12}$~cm for $T \sim 5 \times 10^7$~K gas) of the outflowing material within the simulation domain. There is some evidence based upon visualizations of the density and magnetic field that our result may be more complex than a pure infall or thermal wind outflow, and may consist of a combination of a thermal wind and reconnection-driven infall \citep {igumenschevetal03}.  The total net mass outflow and infall is, of course, central to the determination of the nuclear energetic yield and corresponding light curves of any possible SNe~Ia that originates from a double-degenerate merger --- not only of sub-Chandrasekhar mass as studied here, but potentially also of super-Chandrasekhar mass if the merger does not result in a prompt detonation. 

Furthermore, the topology of the seed field plays a crucial role in determining whether a large-scale ordered field gives rise to jets \citep {beckwithetal08}. Consequently, our findings with regard to magnetic braking, as well as the jet and outflow, are to some extent dependent upon our initial seed magnetic field, which consists of a single torus of magnetic flux. Further work will need to be done to determine the degree to which these conclusions are robust in the presence of realistic seed magnetic fields generated from the white dwarf merger process itself.

\section{Conclusions}
\label{sec:discussion}

The nuclear burning time at the endpoint of our standard simulation $t_{\rm CC} \sim 1$~yr is much longer than our evolutionary time of 6 hours. Thus, while the central temperature continues to rise even throughout the simulation, the outcome of the nuclear burning is not yet clear. The conditions are, however, favorable  to supersonic detonations, as opposed to subsonic deflagrations. At a density $\sim 10^7$~g~cm$^{-3}$, the specific energy release in carbon burning exceeds the specific internal energy by about an order of magnitude, which gives rise in an overpressure also about an order of magnitude greater than the background pressure --- see, for instance, \cite {nomoto82}. Thus our results suggest that under appropriate conditions of near-equal mass non-synchronous mergers, a sub-Chakdrasekhar merger may give rise to a central detonation powering a SNe~Ia. However, while this scenario is promising, a detonation is not assured, as the detonation initiation condition depends on the temperature profile \citep {seitenzahletal09a}. Consequently, further computational studies will need to be carried out to explore the outcome in greater detail. 

Additionally, while the 2D axisymmetric models presented here begin to shed light on the role of the magnetic field in binary white dwarf mergers, both the accretion itself and the development of the Maxwell stresses leading to magnetic braking of the rotating white dwarf are necessarily limited in 2D axisymmetry. Additionally, some of the long-lived magnetic flux tori which we see in 2D may or may not prove to be stable in 3D, which may significantly alter the evolution of the corona. A fuller picture of the evolution of the rotating white dwarf merger will require 3D simulations, and while these will be demanding, we expect that these will should be more favorable to thermonuclear runaway.

An off-centered ignition, possibly leading to an accretion-induced collapse, may still be possible for significantly unequal mass mergers, but we expect that the influence of the magnetic field  will alter some conclusions of previous models quantitatively, if not qualitatively. Furthermore, we expect that for unequal mass or significantly lower-mass mergers, the peak temperatures reached even during the post-merger phase may never reach ignition conditions. In these cases, the outcome will not be either a SNe~Ia or an accretion-induced collapse, but rather a stable object. Our findings are consistent with theoretical models which have predicted the growth of the white dwarf magnetic field through binary systems \citep {Toutetal08, nordhausetal11, garciaberroetal12}.   Our simulations demonstrate that such an object will be strongly-magnetized, and may account for the existence of HFMWDs in general and hot DQ white dwarfs specifically. Moreover, if both the double-degenerate channel of white dwarf mergers, as conjectured by \citet {garciaberroetal12}, and a separate channel of white dwarf-low mass companions, as conjectured by \citet {nordhausetal11},  are simultaneously active, we expect the HFMWD distribution to be bimodal in field strength. 

HFMWDs and SNe~Ia may therefore represent disparate outcomes of the same fundamental astrophysical process of merging white dwarfs. Consequently, the HMWWD magnetic field distribution may help inform our understanding of which double-degenerate systems  are failed SNe Ia, yielding stable white dwarfs as opposed to thermonuclear detonations. Further magnetized models of a wide range of white dwarf masses, ranging from ONe white dwarfs to He white dwarfs through low-mass stellar and planetary companions,  will help establish the HMFWD birth field distribution, which can then be modeled to compare directly against their observed values in the field \citep {vanlandinghametal05}. Such a study will also help more fully elucidate the conditions for thermonuclear runaway to SNe~Ia in the double-degenerate channel.

As we have demonstrated, the magnetorotationally-driven disk turbulence produces a outflux of $\sim 10^{-3} M_{\odot}$ of unbound mass from the disk. We expect such magnetically-driven outfluxes to be a general outcome of the MRI during the post-merger stage of evolution of the double-degenerate system. Therefore, our models suggest that the immediate CSM environment surrounding double degenerate white dwarf mergers may not be as clean as previously believed. This finding has significant ramifications for observational studies of the CSM surrounding SNe Ia using NaID absorption lines in the optical, as well as in the radio and X-ray. In particular, if a delay of $\sim 10^4$ yr follows the initial white dwarf merger, the cooled magnetized outflows may ultimately give rise to NaID lines though the snowplow effect as the outflow interacts with the interstellar medium. However, given the dynamics of the snowplow and the low densities of the ISM, it may be challenging to account for the existence of CSM nearby $\sim 10^{16}$ cm some SNe Ia \citep {patatetal07} in the context of the double-degenerate channel. Our simulations demonstrate these magnetized outflows to be strongly asymmetric, with an opening angle of $\sim50 ^{\circ}$. Furthermore, to date, only upper limits for the X-ray luminosity  for both 53 SNe~Ia using Swift \citep {russellimmler12}, as well in the X-ray and radio for 2011fe \citep {horeshetal12, chomiuketal12} have been established.  Further magnetized models on extended domain sizes, using a range of seed fields, and capturing both the explosion and the sonic point of the outflow material may help elucidate a precise, predictive observational signature of magnetized outflows from the double-degenerate channel of SNe~Ia. Such simulations will address the final fate of the outflows --- a wind, infall, or a combination thereof. They will also address to what extent if any the outfluxes may have on the light curve and spectra at early times, as the supernova blast wave encounters outfluxed circumstellar matter surrounding the double-degenerate system, or infall back onto the white dwarf merger powers the light curve through accretion energy.

The results presented here are a first step towards a fuller understanding of the post-merger evolution of the coalescence of binary white dwarfs. Clearly, more work in this direction is needed, including full 3D MHD simulations. We expect that just as with many other important astrophysical systems, including both core collapse and single-degenerate models of SNe Ia, 2D simulations will help shape our understanding of the key physical processes involved in double-degenerate mergers. Yet, the limitations of 2D ultimately require us to make the leap to full 3D studies, and  work in this direction is already in progress.  Dimensionality is a crucial issue, since the development of the magnetic field and turbulence, and possibly the observational properties of the merger remnant may be qualitatively different when simulated in 3D. The current work paves the road towards a fuller understanding of the magnetized merger remnants.


\appendix 
\label {appendix}

In this appendix, we detail the azimuthal-averaging procedure used to average the 3D SPH particles onto an axisymmetric Eulerian $r$-$z$ mesh. A simple method involves averaging the 3D SPH particles onto a 3D cylindrical Eulerian mesh, then angle-averaging this mesh. However, a more elegant and efficient method, which we adopt, is to break up the angle-dependent quantities, which involve only the smoothing kernel, from the particle data. This reduces the angle-averaging procedure to simple quadrature, with the weights written as integrals to be pre-tabulated once. The averaged quantities are then rapidly and efficiently calculated by weighting the particle data at each point on the $r$-$z$ mesh over lookups of the pre-tabulated weights.

The kernel function is the central mathematical object within the context of SPH which allows continuous fluid quantities to be calculated from discrete particle data. We adopt the form \citep {monaghanlattanzio85}:

\begin {equation}	
 W(\mathbf{r}, h) = \frac{1}{\pi\,h^3} \begin{cases}
2\,(\frac{3}{4}-\nu^2), 	& 0\le\nu\le\frac{1}{2}\\
(\frac{3}{2}-\nu)^2,	 	& \frac{1}{2}<\nu\le\frac{3}{2}\\
0,						 	& \nu>\frac{3}{2}\end{cases} 
\end {equation}
Here  $\nu$ is the dimensionless distance scaled to the smoothing length --- $\nu={|\mathbf{r}|} / h$. We further define a rescaled dimensionless kernel function $\tilde {W}$ as 
\begin {equation}
\tilde{W} (\nu) = \frac{4}{3}\,\pi\,h^3\  W(\mathbf{r},h)= \begin{cases}
\frac{8}{3}\,(\frac{3}{4}-\nu^2),\	& 0\le\nu\le\frac{1}{2}\\
\frac{4}{3}\,(\frac{3}{2}-\nu)^2,\ 	& \frac{1}{2}<\nu\le\frac{3}{2}\\
0,\ 								& \nu>\frac{3}{2}\end{cases} 
\end {equation}
$\tilde{W}$ depends solely on $\nu$. In the ``scatter'' interpretation of SPH, each quantity $A$ at a specific position $\mathbf{r}$ is given by
\begin {equation}
A(\mathbf{r}) = \sum_{i=1}^{N} A_i\,\frac{m_i}{\rho_i}\,W(|\mathbf{r}-\mathbf{r_i}|, h_i),
\end {equation}
with the number of particles $N$ and subscripts denoting quantities for a specific particle. To angle-average this quantity $A$, we integrate it over the angle $\varphi$ and divide it by $2\,\pi$:
\begin {equation}
 A(r,z) = \frac{1}{2\,\pi}\int_0^{2\pi} \text{d}\varphi\,A(r,\varphi,z) 
\end {equation}
Because the quantities $m_i$, $\rho_i$ and $A_i$ do not depend upon angle, we may interchange the order of integration and summation, and write $A(r, z)$ as
\begin{equation}
A(r, z) = \sum_{i=1}^{N} A_i\,\frac{m_i}{\rho_i}\,\frac{1}{2\pi}\int_0^{2\pi} \text{d}\varphi\,\frac{1}{\frac{4}{3}\,\pi\,h_i^3}\,\tilde{W}(\nu).
\label{average}
\end{equation}
%
%
%
Written in this way, we can transparently see that in the scatter interpretation, angle-averaging involves only the smoothing kernel itself. We can further simplify this expression by defining the dimensionless  2D cylindrical distance $\nu_{\rm 2D}= \sqrt{(r-r_i)^2+(z-z_i)^2} /h_i$ and  dimensionless ratio $x  = 2 r r_i / h_i^2$  to express

%
%

%
\begin {equation}
\nu  =  \sqrt{{\nu_{\rm 2D}}^2 + x\,[1-\cos(\varphi-\varphi_i)]} 
\end {equation}
%
Next, we define $W_0(\nu_{\rm 2D},x)$ to be the angle-averaged dimensionless kernel, expressed as a function of $\nu_{\rm 2D}$ and $x$:
\begin{equation}
W_0(\nu_{\rm 2D},x)  = \frac{3}{8\pi^2}\int_0^{2\pi}\text{d}\varphi\,\tilde{W} (\nu_{\rm 2D},x,\cos(\varphi)) 
\end{equation}
This allows us to efficiently compute the azimuthally-averaged temperature $T$ and the z-velocity $v_z$ at each point as
\begin {equation}
\rho(r, z) = \sum_{i=1}^N \frac{m_i}{h_i^3 }\,W_0 (\nu_{\rm 2D},x) \text{,}\qquad   T(r, z) = \sum_{i=1}^N T_i \frac{m_i}{\rho_i h_i^3}\,W_0 (\nu_{\rm 2D},x) \text{,} \qquad  \text{and}\qquad v_z (r, z) = \sum_{i=1}^N v_{z, i} {m_i \over \rho_i h_i^3}\,W_0 (\nu_{\rm 2D},x).\
\end {equation}

One can similarly show that the $r$ and $\phi$ velocity components can be expressed as
\begin {equation}
 v_r (r,z) = \sum_{i=1}^N \left[v_{r,i} {m_i \over \rho_i h_i^3}\,W_{0,\cos}(\nu_{\rm 2D},x) + v_{\varphi,i} {m_i \over \rho_i h_i^3}\,W_{0,\sin}(\nu_{\rm 2D},x) \right],
\end {equation}
and
\begin {equation}
v_\varphi (r,z) = \sum_{i=1}^N \left[ v_{\varphi,i}\,{m_i \over \rho_i h_i^3}\,W_{0,\cos}(\nu_{\rm 2D},x) + v_{r,i}\,{m_i \over \rho_i h_i^3}\,W_{0,\sin}(\nu_{\rm 2D},x) \right] ,
\end {equation}
where we have defined 
\begin {equation}
W_{0,\cos} (\nu_{\rm 2D}, x) = \frac{3}{8\pi^2} \int_0^{2\pi} \text{d}\varphi\,\tilde{W}(\nu_{\rm 2D},x,\varphi)\,\cos \varphi 
\end {equation}
and
\begin {equation}
W_{0,\sin} (\nu_{\rm 2D}, x) = \frac{3}{8\pi^2} \int_0^{2\pi} \text{d}\varphi\,\tilde{W}(\nu_{\rm 2D},x,\varphi)\,\sin \varphi 
\end {equation}
analogously to $W_0 (\nu_{\rm 2D},x)$. In practice, the functions  $W_0 (\nu_{\rm 2D},x)$, $W_{0,\cos} (\nu_{\rm 2D}, x)$, and $W_{0,\sin} (\nu_{\rm 2D}, x)$ can be pre-tabulated as functions of both $\nu_{\rm 2D}$ and $x$. In this way, one can rapidly and efficiently calculate Eulerian grid quantities from azimuthally-averaged SPH data.

\acknowledgements
This research was partially supported by a grant from the American Astronomical Society. Simulations at UMass Dartmouth were performed on a computer cluster supported by NSF grant CNS-0959382 and AFOSR DURIP grant FA9550-10-1-0354. PT, GJ, and DL acknowledge support  at the University of Chicago in part by the U.S Department of Energy (DOE) under Contract B523820 to the ASC Alliances Center for Astrophysical Nuclear Flashes, and in part by the National Science Foundation under Grant No. AST - 0909132 for the ``Petascale Computing of Thermonuclear Supernova Explosions.''    EGB and PLA acknowledge support by MCINN grant  AYA2011--23102, by the European  Union FEDER  funds, by the AGAUR and by  the ESF  EUROGENESIS project (grant EUI2009-04167). Additional simulations at the University of Chicago Beagle cluster are supported by the NIH through resources provided by the Computation Institute and the Biological Sciences Division of the University of Chicago and Argonne National Laboratory, under grant S10 RR029030-01.  This research has made use of NASA's Astrophysics Data System and the yt astrophysics analysis software suite \citep {turketal11}.

\bibliography{fisher_sne}
\bibliographystyle{apj}

\end{document}